\documentclass[aps]{revtex4} 

\usepackage{epsfig,amssymb,bm,graphicx,color,amsmath,rotating}
\usepackage[percent]{overpic}
\usepackage{mwe,xcolor}
\usepackage{transparent}
\usepackage{amsmath}
\usepackage{framed}
\usepackage{slashed}
\usepackage{MnSymbol}

\newcommand{\landau}{\mathcal{O}}
\newcommand{\blau}[1]{\textcolor{black}{#1}}

\begin{document}

\title{Exotic Cores with and without Dark-Matter Admixtures \\ in Compact Stars} 

\author{R.~Z\"ollner, B.~K\"ampfer}
\affiliation{Helmholtz-Zentrum  Dresden-Rossendorf, 01314 Dresden, Germany}
\affiliation{Institut f\"ur Theoretische Physik, TU~Dresden, 01062 Dresden, Germany}

% Abstract (Do not insert blank lines, i.e. \\) 
\begin{abstract}
	We parameterize the core of compact spherical star configurations by a mass ($m_x$) and a radius ($r_x$) and study the resulting admissible areas in the total-mass -- total-radius plane. The employed fiducial equation-of-state models of the corona at radii $r > r_x$ and pressures $p \le p_x$ with $p(r = r_x) = p_x$ are that of constant sound velocity and a proxy of DY$\Delta$ DD-ME2 provided by Buchdahl's exactly solvable ansatz. The core ($r < r_x$) may contain any type of material, e.g.\ Standard-Model matter with unspecified equation of state or/and an unspecified Dark-Matter admixture. Employing a toy model for the cool equation of state with first-order phase transition we discuss also the mass-radius relation of compact stars with an admixture of Dark Matter in a Mirror-World scenario.
\end{abstract}
% Keywords

\date{\today}

\maketitle

\section{Introduction} \label{introduction}

The advent of detecting gravitational waves from merging neutron stars, the related multimessenger astrophysics \cite{Pang:2021jta,Annala:2021gom,Yu:2021nvx,Nicholl:2021rcr,Margutti:2020xbo,Tang:2020koz,Tews:2020ylw,Silva:2020acr}
and the improving mass-radius determinations of neutron stars, in particular by NICER data \cite{Riley:2021pdl,Miller:2021qha,Miller:2019cac}, stimulated a wealth of activities. Besides mass and radius, moments of inertia and tidal deformarbilities become experimentally accessible and can be confronted with theoretical models 
\cite{Chatziioannou:2020pqz,Christian:2019qer,Motta:2022nlj,Jokela:2021vwy,Kovensky:2021kzl,Zhang:2021xdt,Pang:2021jta,Pereira:2020cmv}.
The baseline of the latter ones is provided by non-rotating, spherically symmetric
cold dense matter configurations. The sequence of white dwarfs (first island of stability)
and neutron stars (second island of stability) and --possibly \cite{Christian:2020xwz}-- 
a third island of stability \cite{Gerlach:1968zz,Kampfer:1981zmq,Kampfer:1981yr}
shows thereby up when going to more compact objects, with details depending sensitively
on the actual equation of state (EoS). 
The quest for a fourth island has been addressed too \cite{Li:2019fqe,Alford:2017qgh}.
"Stability" means here the damping of radial disturbances. Since the radii of configurations
of the second (neutron stars) and third (hypothetical quark/hybrid stars 
\cite{Malfatti:2020onm,Pereira:2022stw,Bejger:2016emu}) 
islands are very similar, 
the notion of twin stars \cite{Glendenning:1998ag,Jakobus:2020nxw}.
has been coined for equal-mass configurations; "masquerade" was another related term \cite{Alford:2004pf}.

The standard modeling of such static compact star configurations is based on 
the Tolman-Oppenheimer-Volkov (TOV) equations
\begin{eqnarray}
	\frac{\mbox{d} p}{\mbox{d} r } &=& 
	- G_N \frac{(e+p) (m + 4 \pi p r^3)}{r^2 (1 - \frac{2 G_N m}{r})}, \label{eq:p_prime} \\
	\frac{\mbox{d} m}{\mbox{d} r} &=& 4 \pi e r^2, \label{eq:m_prime}
\end{eqnarray}  
emerging from the energy-momentum tensor of a perfect isotropical fluid and spherical symmetry
of space-time and matter as well, within the framework of Einstein gravity without cosmological term.
Newton's constant is denoted by $G_N$, \blau{natural units with $c = 1$ are used,
unless when relating mass and length and energy density, where $\hbar c$ is needed.}
Given a unique relationship of pressure $p$ and energy density $e$ as equation of state (EoS)
$e(p)$, in particular at zero temperature, the TOV equations are integrated with boundary conditions
\blau{$p(0) = p_c $ and $m(0) = 0$
(implying $p(r) = p_c - \landau (r^2)$ and $m(r) = 0 + \landau (r^3)$} at small radii $r$),
and $p(R) = 0$ and $m(R) = M$ with $R$ as circumferential radius and $M$ as gravitational mass
(acting as parameter in the external (vacuum) Schwarzschild solution at $r > R$).
The quantity $p_c$ is the central pressure.
The solutions $R(p_c)$ and $M(p_c)$
provide the mass-radius relation in parametric form $M(R)$.

A big deal of efforts is presently concerned about the EoS at supra-nuclear densities \cite{Reed:2021nqk}.
Figure 1 in \cite{Annala:2019puf} exhibits the currently admitted uncertainty: up to a factor of ten in pressure 
as a function of energy density. At asymptotically large energy density, perturbative QCD constraints
the EoS, though it is just the non-asymptotic supra-nuclear density region which determines crucially
whether twin stars may exist or quark-matter cores appear in neutron stars. 
Accordingly, one can fill this gap by a big number of test EoSs 
to scan through the possibly resulting mass-radius curves, see 
\cite{Ayriyan:2021prr,Greif:2020pju,Lattimer:2015nhk,LopeOter:2019pcq}.
However, the option that neutron stars can accommodate Dark Matter (or other exotic material)
\cite{Anzuini:2021lnv,Bell:2019pyc,Das:2021hnk,Das:2021yny}
obscures the safe theoretical modeling of a reliable mass-radius relation in such manner. Of course, 
inverting the posed problem with sufficiently precise data of masses and radii as input offers 
a promising avenue towards determining the EoS 
\cite{Newton:2021yru,Huth:2021bsp,Ayriyan:2021prr,Blaschke:2020qqj,Raaijmakers:2021uju,Raaijmakers:2019dks,Raaijmakers:2019qny}.

Here, we pursue another perspective: we parameterize the supra-nuclear core by a radius~$r_x$ and the
included mass~$m_x$ and integrate the above TOV equations only in the corona, i.e.\ from pressure $p_x$
to the surface, where $p = 0$. This yields the total mass $M(r_x, m_x; p_x)$ and the total radius $R(r_x, m_x; p_x)$
by assuming that the corona EoS $e(p)$ is reliably known at $p \le p_x$. 
(Our notion ``corona" could be substituted by ``mantel" or ``crust" or ``envelope".
It refers to the complete part of the compact star outside the core, $r_x \le r \le R$.)
Clearly, without knowledge of the matter composition at $p > p_x$ (may it be Standard-Model matter with uncertain EoS or may it contain a Dark-Matter admixture, for instance, or monopoles or some other type of
"exotic" matter) one does not get  a simple mass-radius relation by such procedure, 
but admissible area(s) over the mass-radius plane, depending on the core
parameters $r_x$ and $m_x$. This is the price of avoiding a special model of the core matter composition.

We are going to test two models of the corona EoS: (i) constant sound velocity with crucial parameter
$e_0 = e(p=0)$ and $p=0$ for $e< e_0$, and (ii) Buchdahl's EoS which, surprisingly, represents a 
proxy of the DY$\Delta$ DD-ME2 EoS \cite{Li:2019fqe} and has the property $\lim_{p \to 0} e(p) \to 0$.
One could consider this as a test of the sensitivity against variations of the neutron star crust
(see \cite{Suleiman:2021hre,Lattimer:2006xb} for such a crust test 
within the traditional approach with given core EoS). 

We emphasize the relation of ultra-relativistic heavy-ion collision physics, related
to the EoS $p(T, \mu_B \approx 0)$, and compact star physics, related to
$p(T \approx 0, \mu_B)$ when focusing on static compact-star properties
\cite{Klahn:2006ir,Most:2022wgo}.
(Of course, in binary or ternary compact-star merging-events, 
also finite temperatures $T$ and a large range of baryon-chemical potential $\mu_B$ are probed 
\cite{HADES:2019auv}.)
Implications of the conjecture of a first-order phase transition at small temperatures and 
large baryo-chemical potentials or densities
\cite{Stephanov:1999zu,Karsch:2001cy,Fukushima:2010bq,Halasz:1998qr}
can be  studied by neutron-hybrid-quark stars \cite{Blaschke:2020vuy,Blacker:2020nlq,Cierniak:2021knt,Orsaria:2019ftf}.
It is known since some time 
\cite{Gerlach:1968zz,Kampfer:1985mre,Kampfer:1983we,Kampfer:1981yr,Kampfer:1981zmq}
that a cold EoS with special pressure-energy density relation $p(e)$, e.g.\ a strong local softening
up to first-order phase transition with a density jump, 
can give rise to a ``third family" of
compact stars, beyond white dwarfs and neutron stars. In special cases, the third-family stars appear
as twins of neutron stars \cite{Schertler:2000xq,Alford:2004pf,Christian:2017jni}.
Various scenarios of the transition dynamics to the denser configuration as
mini-supernova have been discussed also quite early \cite{Migdal:1979je,Kampfer:1983zz}. 

Our paper is organized as follows. In Section \ref{csv} we employ the corona EoS model (i) 
(constant sound velocity)
and consider also the instructive limiting case of an incompressible fluid. 
Special relations of $m_x (r_x)$ are deployed to emphasize the relation to the traditional approach,
which emerges as cut in the $M$-$R$ plane representing the curve $M(R)$. 
Section \ref{NYD} uses the corona EoS model (ii)
(Buchdahl's EoS \blau{as proxy of a nuclear-physics based EoS.} The $M(R)$ relations for models (i) and (ii)
are quite different, as the admitted areas over the $M$-$R$  plane are too.  
We conclude in section \ref{Summary}.
Appendix \ref{DM_NS} considers Dark-Matter or Mirror-World admixtures 
in neutron (quark) stars with a phase transition with the goal 
of illustrating the potential impact of a first-order phase transition on the mass-radius relation
and the quest for twin stars. This is another way to specify the relation $m_x (r_x)$.

\section{Corona with constant sound velocity EoS} \label{csv}

The constant sound velocity EoS
\begin{equation} \label{eq:csvEoS}
	p(e) = v_s^2 (e - e_0) \,\, \mbox{for} \, e > e_0, \quad p = 0  \,\, \mbox{for} \, e \le e_0
\end{equation}
introduces a scale setting parameter $e_0$ which drops out in the representation by
scaled quantities: $\bar p = p / e_0$, $\bar r = r \sqrt{G_N e_0}$, $\bar m = m G_N \sqrt{G_N e_0}$.
Only the sound velocity $v_s^2 = \partial p / \partial e$ remains as parameter. We use this as model (i)
and take --arbitrarily-- $\bar p_x = 1$  as upper limit of scaled pressure of the corona EoS.

\subsection{A limiting case: infinite sound veleocity}\label{Schwarz}

\begin{figure}[tb!]
	\includegraphics[width=0.88\columnwidth]{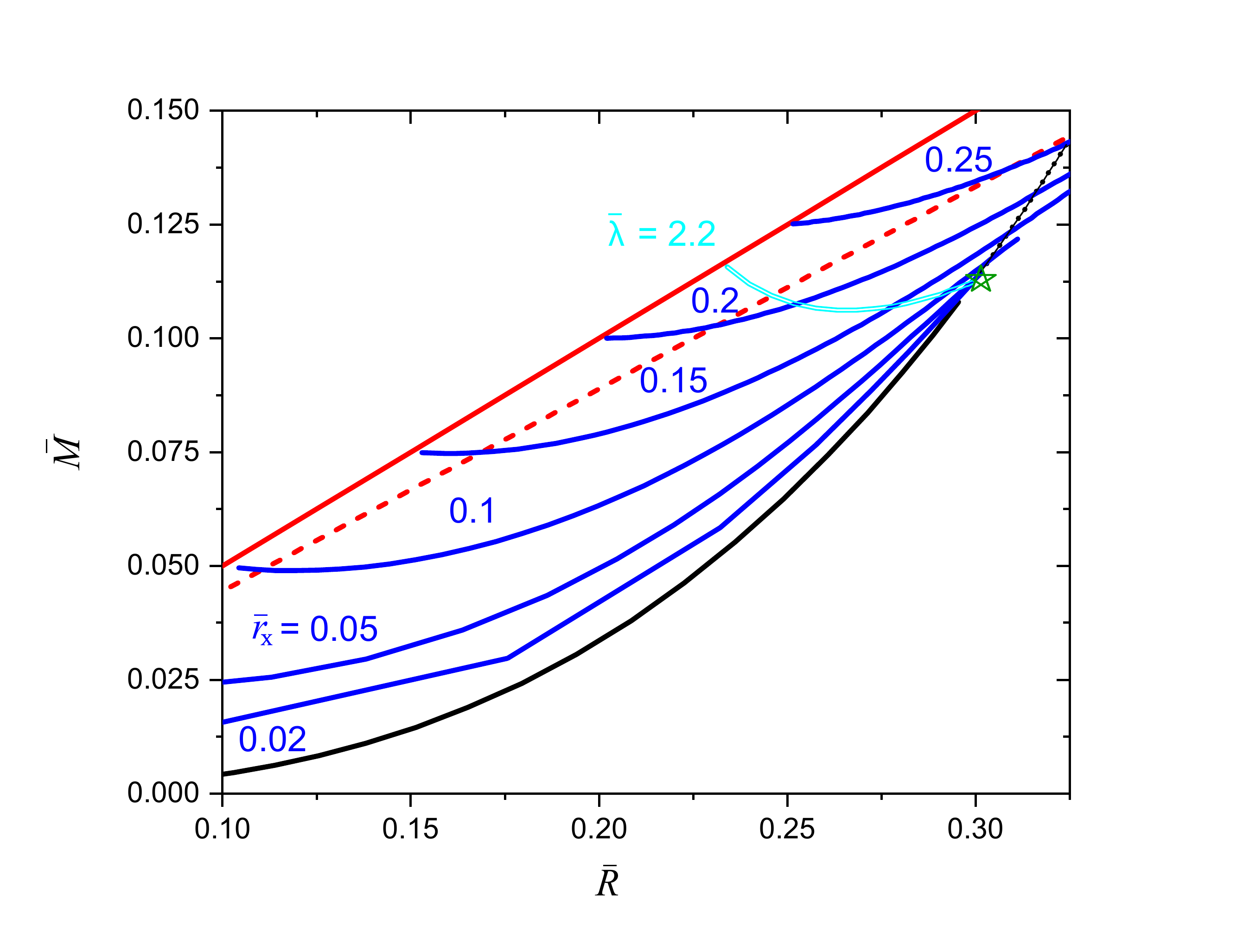}
	\caption{Admissible area in the $\bar M$-$\bar R$ plane mapped out by curves $\bar r_x = const$ \blau{(blue solid, obtained by integrating the TOV equations from $\bar r_x$ to $\bar R$)}.
		The black hole condition $\bar M = \bar R / 2$ is depicted by the red fat line. Buchdahl's limit
		$\bar M = 4 \bar R / 9$ is displayed by the dashed red line.
		The Schwarzschild curve $\bar M = \frac{4 \pi}{3} \bar R^3$ (black curve) terminates
		in the asterisk $\pentagram$, where $\bar p_c =1$ is reached. Allowing for a larger central pressure, continues the
		Schwarzschild curve (dotted black curve), which coincides with the core-corona model for
		$\bar m_x (\bar r_x) = \frac{4 \pi }{3} \bar r_x^3$.
		Imposing the side condition $\bar m_x (\bar r_x) = \frac{4 \pi }{3} \bar \lambda \bar r_x^3$
		results for $\bar \lambda = 2.2$ in the cyan double curve. 
		The actual calculations are for Eqs.~(\ref{eq:p_prime}, \ref{eq:m_prime}, \ref{eq:csvEoS})
		with $v_s^{-2} = 10^{-4}$, \blau{to approximate an EoS with constant energy density}.
		\label{fig:1} 
	}
\end{figure}

It is instructive to consider the limiting case $v_s^{-2} \to 0$, i.e.\ \blau{$\bar e \to 1$} for $\bar p \le 1$.
The constant energy density in Eq.~(\ref{eq:m_prime}) results in 
$\bar m = \bar m_x + \frac{4 \pi}{3} (\bar r^3 - \bar r_x^3)$, and the \blau{Riccati type ordinary differential equation} (\ref{eq:p_prime})
becomes\blau{, by a suitable shift of the pressure,} a Bernoulli equation allowing a quadrature to get the pressure profile 
$\bar p (\bar r) = - 1 + (\bar p_c + 1) \exp\{ \frac12 (\lambda - \lambda_0\} {N}$,
${N}^{-1} = 1 + 4 \pi  (\bar p_c + 1) \exp\{ - \frac12 \lambda_0 \}
\int_{\bar r_x}^{\bar r} \mbox{d} \tilde r \, \tilde r \exp\{ \frac32 \lambda (\tilde r) \}$,
$e^{- \lambda} = 1 - 2 \bar m / \bar r$, $\lambda_0 = \lambda (\bar r_x)$,
cf.\ \cite{Kampfer:1981zmq,Haensel:1987}. 
Since the term with $\int_{\bar r_x}^{\bar r} \mbox{d} \tilde r \tilde r \exp\{ \frac32 \lambda (\tilde r) \}$
is less transparent due to the elliptic integral(s), we \blau{turn on an approximation of the case $e = const$ and} integrate numerically Eqs.~(\ref{eq:p_prime}, \ref{eq:m_prime})
with $v_s^{-2} = 10^{-4}$ \blau{to mimic the above constant energy density EoS by a proxy with large sound velocity. We then} find admissible area in the $\bar M$-$\bar R$ plane displayed in Fig.~\ref{fig:1}.
That area is mapped out \blau{here} by curves $\bar r_x = const$ with $\bar m_x$ varying from small (at the r.h.s.)
to large (at the l.h.s.) values. The l.h.s.\ limitation is given by the black hole condition $\bar M = \bar R / 2$
(red fat line). In the limit of small cores, the $\bar r_x = const$ curves 
$\bar M = \bar m_x + \frac{4 \pi}{3} \bar R^3 - {\landau}(r_x^3)$
approach the Schwarzschild mass-radius relation $\bar M = \frac{4 \pi}{3} \bar R^3$ (dashed black curve),
i.e.\  the gap to the $\bar r_x = const$ curve becomes larger with decreasing values of $\bar R$ due to
increasing $\bar m_x$. The Schwarzschild curve terminates in the asterisk since there  
$\bar p_c =1$ is reached. One may recall the pressure profile of the interior Schwarzschild solution
$\bar p_{Schwarz} (\bar r) = - [(1+\bar p_c) - (1 + 3 \bar p_c) W] /[3(1+\bar p_c) - (1 + 3 \bar p_c) W]$,
$W := \sqrt{1 - \frac{8 \pi}{3} \bar r^2}$,
yielding $\bar R (\bar p_c)$ from $\bar p (\bar R) = 0$
to see this via our proposition $\bar p \le \bar p_x$.

In this context, it is interesting to compare the pressure profiles of the interior Schwarzschild solution
and the core-corona model at small values of $\bar r_x$. Adjusting the central pressure of the
interior Schwarzschild solution such to obtain the same scaled radius $\bar R$ as in the core-corona model
with given small value of $\bar r_x$ and given value of $\bar m_x$, one finds (not displayed)
a rapidly dropping pressure in a narrow region above $\bar r_x$ towards the Schwarzschild pressure
$\bar p_{Schwarz} (\bar r)$ which holds until the surface. In such a way, the admissible area of masses and
radii of the core-corona model is bracketed by the black hole limit at l.h.s.\ and the conventional 
mass-radius curve in the lower part of the mass-radius plane.

Let us now consider two possible sections through the mass-radius plane. First, we assume a core-mass
vs. core-radius relation as $\bar m_x (\bar r_x) = \frac{4 \pi }{3} \bar r_x^3$. This yields the dotted curve
which, clearly, coincides with the Schwarzschild relation when continuing the latter one to central pressures
larger than $\bar p_x$. 
Note that with increasing values of $\bar r_x$ the admissible region of the core-corona model
leaks increasingly into the r.h.s.  

A second side condition in relating the core-mass and core-radius is given by
$\bar m_x (\bar r_x) = \frac{4 \pi }{3} \bar \lambda \bar r_x^3$, meaning that
the energy density in the core is $\bar \lambda$ times the energy density of the corona.
(This is essentially the two-shell model considered in \cite{Kampfer:1981zmq,Kampfer:1981yr}). 
Selecting $\bar \lambda = 2.2$
facilitates the cyan double curve, again a well defined mass-radius curve. Larger (smaller) values
of $\bar \lambda$ bend down (up) that curve. The section between the local r.h.s.\ \blau{mass} maximum \blau{at the asterisk} and
the local \blau{mass} minimum belongs to unstable configurations, while left to the local \blau{mass} minimum the
configurations become stable again. This very construction is at the heart of twin stars, i.e.\
stable configurations with the same mass but (slightly) different radii.
Notable is the leakage into the region l.h.s.\ to the Buchdahl bound 
(cf.~\cite{Schaffner-Bielich:2020psc} for a corollary)
given by $\bar M = 4 \bar R / 9$.
This issue is caused by the fact that a central pressure $\bar p_c < \infty$ does not drop to $\bar p_x$
at radius $\bar r_x$ when using the TOV equations with core density $\bar \lambda \bar e_0$.
Phrased differently, the above {\it ad hoc} choice $\bar m_x (\bar r_x) = \frac{4 \pi }{3} \bar \lambda \bar r_x^3$
is compatible with the TOV equations only for a limited range of values of $\bar r_x$. The validity of the TOV equations
for a consistent (always decreasing) pressure profile from $p_c$ to zero limits the extension of the
curve under discussion and lets it terminate before reaching Buchdahl's limit. Nevertheless, extensions of the
TOV equations, e.g.\ as system describing a multi-fluid medium with Dark-Matter components,
(see Appendix \ref{DM_NS})
make the occupancy of the region beyond Buchdahl's limit conceivable.   

\subsection{Finite sound velocity}\label{fsv}

\begin{figure}[tb!]
	\includegraphics[width=0.88\columnwidth]{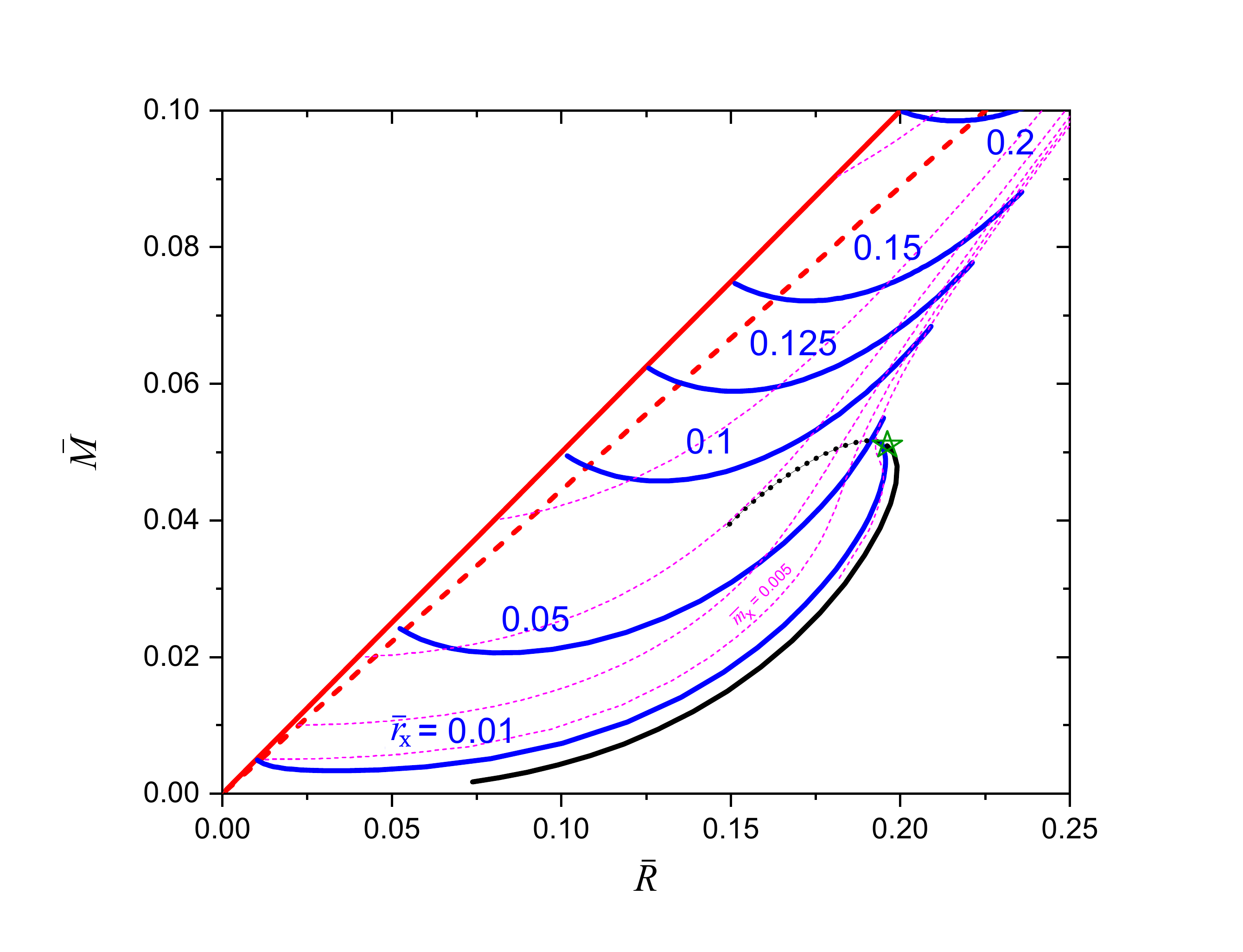}
	\caption{As Fig.~\ref{fig:1} but for $v_s^{-2} = 3$ \blau{to cope with causality}.
		The dashed magenta curves depict \blau{additionally} mass-radius relations for running values of $\bar r_x$  
		and $\bar m_x = const$ with values of 0.01, 0.02, 0.04, and 0.09 above the labeled $\bar m_x = 0.005$ curve; 
		the rightmost curve for  $\bar m_x = 0.001$ is displayed only for $\bar M > 0.03$.
		\label{fig:2} 
	}
\end{figure}

For a finite sound velocity, the mass-radius curve bends to left and facilitates a mass maximum,
e.g.\ $\bar M_{max} = 0.051$ at radius $\bar R = 0.195$ for $v_s^{-2} = 3$.
By the requirement of $M \approx 2 M_\odot$ one gets $e_0 \approx 231.5$~MeV/fm${}^3$ and finds, from
$R = M G_N \bar R / \bar M$, a radius of \blau{11.27~km, (accidentally) in the right ballpark.}

The mass-radius plane is exhibited in Fig.~\ref{fig:2} with the same line style conventions as in
Fig.~\ref{fig:1}. One meets the features discussed in the previous Subsection. The admissible region
is bracketed l.h.s.\ by the black hole condition and r.h.s.\ by the plain mass-radius curve $\bar M (\bar R)$
(dashed black curve) up to the asterisk, where $\bar p_c = \bar p_x$ is reached. In the dotted section,
the central pressures obey $\bar p_c > \bar p_x$. Due to non-zero matter compressibility, the analog
relation $\bar m_x (\bar r_x)$, discussed in Subsection \ref{Schwarz}, must be constructed numerically
to see that this side condition within the core-corona model coincides in fact with the mass-radius curve
obtained by the ({\it ad hoc}) continuation of the EoS (\ref{eq:csvEoS}) at $p > p_x$.
For small values of $\bar r_x = const$, the permitted masses as a function of radius approach the 
mass-radius curve $\bar M (\bar R)$. For larger constant values of $\bar r_x$, the pattern of the curves
$\bar M (\bar R, \bar r_x, \bar m_x; \bar p_x)$ is as discussed in Subsection \ref{Schwarz}.
They leak also beyond the Buchdahl limit.

We relegate the consideration of special cuts $\bar m_x (\bar r_x)$ \blau{and variations of $p_x$} to the Appendix \ref{DM_NS}, where we include explicitly an admixture of Dark Matter to be dealt within a two-fluid approach. 

\section{Corona with Buchdahl's EoS and NY$\Delta$ DD-EM2}\label{NYD}

\begin{figure}[t!]
	\includegraphics[width=0.88\columnwidth]{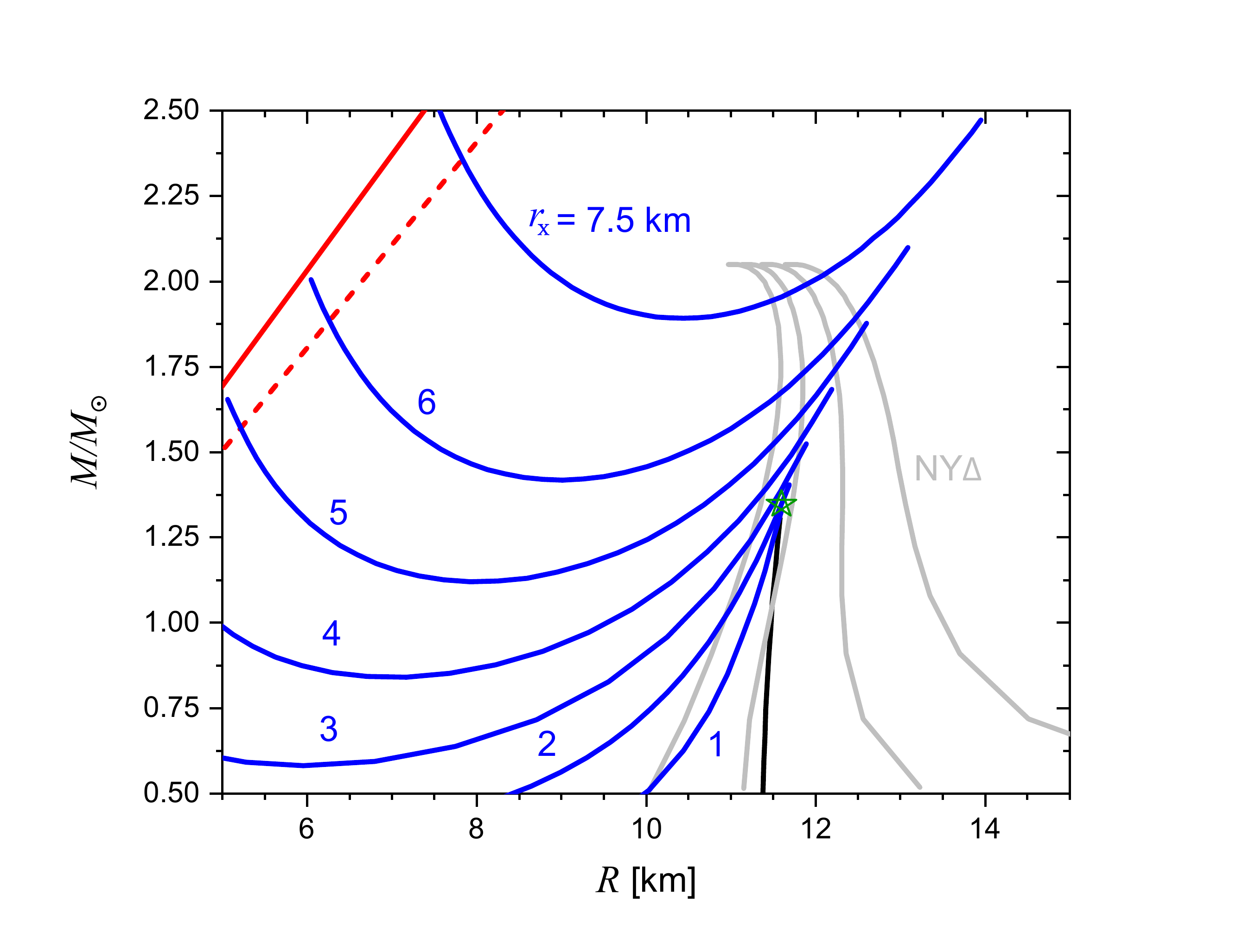}
	\caption{As Fig.~\ref{fig:1}, however for the corona model (ii), 
		i.e.\ EoS (\ref{BuchEoS}) with $p_* = 64$~MeV/fm${}^3$ (solid black curve).
		Note the different units.
		Faint gray curves are based on the \blau{standard integration of the TOV equations by using the} numbers of the NY$\Delta$ EoS tabulated in \cite{Li:2019fqe}
		with linear interpolation both in between the mesh and from the tabulated minimum energy density to the
		$p=0$ point at $e_0$. The chosen values are $e_0 =100,$ 10, 1 and 0.1~MeV/fm${}^3$
		(from left to right). 
		The variation of the low-mass sections of these mass-radius curves points to
		some sensitivity on the outer crust properties.
		The slight deviation of the solid black curve and the second (from left) gray curve points 
		to a non-perfect match of (\ref{BuchEoS}) and NY$\Delta$.
		\label{fig:3} 
	}
\end{figure}

While the employed corona EoS in Section \ref{fsv} displays a finite positive slope of the mass-radius curve
for a large range, the currently preferred EoS models point to a steep increase. An example of an EoS
compatible with neutron star data is NY$\Delta$ based on the DD-ME2 density functional \cite{Li:2019fqe}.
Surprisingly, Buchdahl's EoS \cite{Buchdahl}
\begin{equation}  \label{BuchEoS}
	e = 12 \sqrt{p_* p} - 5 p, \quad p \le p_*
\end{equation}
with $p_* \approx 64$~MeV/fm${}^3$ matches fairly well  NY$\Delta$ within the range tabulated in \cite{Li:2019fqe}, see Table I there.
Due to causality requirement \blau{$v_s^2 \leq 1$}, Eq.~(\ref{BuchEoS}) is limited to  $p \le p_*$. 
In contrast to the EoS of model (i), Eq.~(\ref{BuchEoS}) 
extends smoothly to $e = 0$ and $p =0$. 
The mass-radius relation is given by $M = \beta R /G_N$, and the radius is parameterized
by $R = \frac{1 - \beta}{\sqrt{1 - 2 \beta}} \sqrt{\frac{\pi}{288 p_* G_N}}$ for $\beta \in [0, 1/6]$
\cite{Buchdahl,Lattimer:2000nx}. 
We \blau{use in this section physical units, e.g. $p_*$ in MeV/fm$^3$ instead of $\bar p_x$, and} display masses in units of $M_\odot$ and radii in units of km. With the above quoted value of $p_*$ one finds
$R = 11.33 \frac{1 - \beta}{\sqrt{1 - 2 \beta}}$~km and $M = 7.66 \beta \frac{1 - \beta}{\sqrt{1 - 2 \beta}}  M_\odot$, where the maximum mass $1.31 M_\odot$ and radius $11.56$~km can be read off.

The resulting mass-radius plot is exhibited in Fig.~\ref{fig:3} with line style conventions as in Fig.~\ref{fig:1}.
All the features discussed in Section \ref{csv} are recovered, e.g.\ the convex shape of the curves
$r_x = const$, the approach of these curves to the mass-radius relation based on the EoS (\ref{BuchEoS})
at small values of $r_x$, the occupancy of a region beyond the Buchdahl limit up to the black hole limit etc.
Since $p_x = p_*$ is comparatively small, the resulting mass-radius curve (solid black curve) terminates
at rather small mass, quoted above. However, the core-corona decomposition shows that the maximum-mass
region of NY$\Delta$-DD-ME2 (see gray curves)
is easily uncovered too, interestingly with sizeable core radii and noticeably smaller
up to larger total radii.

\section{Summary} \label{Summary}

We investigate a scenario which assumes \blau{the knowledge of} a trustable equation of state (EoS) for compact spherical stars at pressures $p \le p_x$. In a core-corona decomposition, the core with \blau{an unspecified composition and thus} unknown 
(maybe, exotic matter) \blau{equation(s)} of state is parameterized by a mass $m_x$ and a radius $r_x$;
the pressure at $r_x$ is just $p_x$.
Besides the unspecified equations of state at $p > p_x$, the core may contain also any type of matter,
most notably Dark Matter.  By definition, outside the core --in the corona-- only 
Standard-Model neutron-star matter is present, described by the \blau{fiducial ("trustable")} EoS.
The solution of the TOV
equations of the corona pressure profile $p(r)$ for radii $r \ge r_x$ delivers a broad range of admissible
masses and radii of the total core-corona configurations, bracketed by the black hole limit 
(thus going beyond the Buchdahl limit) and, in part, by the conventional mass-radius curve. 
We explore these admissible region \blau{by using} two test EoS models of the corona \blau{as simple substitutes of a state-of-the-art EoS}: one with a constant sound velocity EoS
and another one which refers to Buchdahl's EoS which in turn approximates \blau{some part of} a particular density functional approach, i.e.\ NY$\Delta$ DD-ME2. The latter one is compatible with current mass-radius observations.

One may imagine that all possible continuations of \blau{a fiducial ("trustable")} EoS beyond $p_x$ generate mass-radius curves
within the admissible region, even with the Buchdahl limit as border line.
However, the admissible region is larger, since it can refer to cores which are not described by the
isotropic one-fluid TOV equations. A straightforward extension of the latter ones is provided
by the two-fluid TOV equations, dealt with in the Appendix.    
In line with investigations of the impact of the cool EoS on compact star properties,
including the intriguing quest for twin stars, we add this schematic consideration of compact-star
mass-radius relations for an EoS with first-order phase transition. In doing so, 
the two-fluid model is deployed where the second fluid is aimed at describing Dark Matter
leaking out of a Mirror World. This scenario facilitates also the option of twin stars for
a wide range of the Dark Matter admixture.

\begin{acknowledgements}
	One of the authors (BK) acknowledges continuous discussions with J.~Schaffner-Bielich and K.~Redlich for the encouragement to deal (again) with the current topic. \\
	The work is supported in part by the European Union’s Horizon 2020 research and innovation program STRONG-2020 under grant agreement No 824093.
\end{acknowledgements}

\begin{appendix}
\section[\appendixname~\thesection]{Dark-Matter/Mirror-World admixtures in neutron stars with phase transition} \label{DM_NS}

Spherical multi-fluid configurations in hydrostatic equilibrium are determined 
by the generalized TOV equations (cf.~\blau{\cite{Dengler:2021qcq, Karkevandi:2021ygv}})
\begin{eqnarray} \label{TOV}
	\frac{\mbox{d} \bar p^{(i)}}{\mbox{d} \bar r} &=& - \left({\bar e}^{(i)} + \bar {p}^{(i)} \right) 
	\frac{1}{r^2}
	\frac{\sum \bar m^{(i)} + 4 \pi \bar r^3 \sum \bar p^{i}}{1 - 2 \frac{\sum \bar m^{(i)}}{\bar r}}, \\
	\frac{\mbox{d} \bar m^{(i)}}{\mbox{d} \bar r} &=& 4 \pi \bar r^2 \bar e^{(i)},
\end{eqnarray}  
where the index $i$ labels the fluid components, which are \blau{here assumed to be} not mutually interacting but are governed by the common gravity field. The quantities are scaled commonly by $e_0$, which figures in the EoS
\begin{equation} \label{EoS}
	\bar p = \left\{ \begin{array}{l}
		0 \quad \mbox{for} \quad  \bar e \le 1 , \\
		v_s^2 (\bar e - 1) \quad \mbox{for} \quad 1 < \bar e < 1 + v_s^{-2} \bar p_x, \\
		\bar p_x \cdots \infty  \quad \mbox{for} \quad \bar e = \bar \lambda (1 + v_s^{-2} \bar p_x) .
	\end{array} \right.
\end{equation}
The incompressible part at $\bar p \ge \bar p_x$ mimics a stiff EoS above the phase transition
at $\bar p_x$ with energy density jump $\bar \lambda$, i.e. $\Delta \bar e = (\bar \lambda - 1) (1+v_s^2 \bar p_x)$.
Phrased differently, the EoS Eq.~(\ref{EoS}) is a toy model of a linear EoS,
$p = v_s^2 (e - e_0)$ for $e \in (e_0, e_x)$, and $p = p_x + \hat v_s^2 (e - \lambda e_x)$ 
for $e \ge \lambda e_x$, where $p_x$ and $e_x$ are related by $p_x = v_s^2 (e_x - e_0)$ \cite{Haensel:1983}.
To reduce the parameter space, we consider $\hat v_s^2 \to \infty$ as limiting stiffness. Due to deploying the scaled quantities, one gets rid of the explicit value of the surface energy density $e_0$, where $p(R) = 0$ holds.

\begin{figure}[t!]
	\includegraphics[width=0.44\columnwidth]{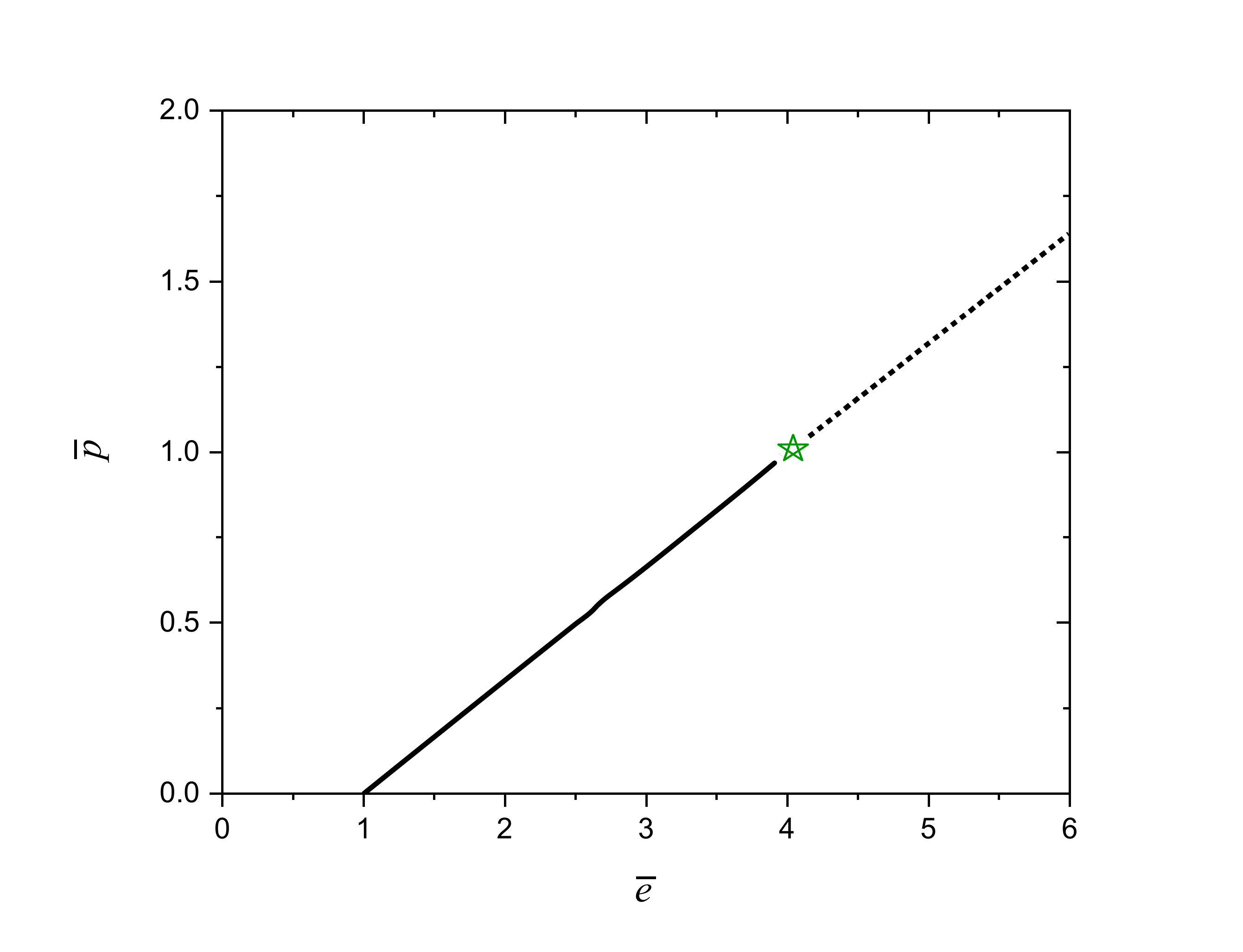}
	\includegraphics[width=0.44\columnwidth]{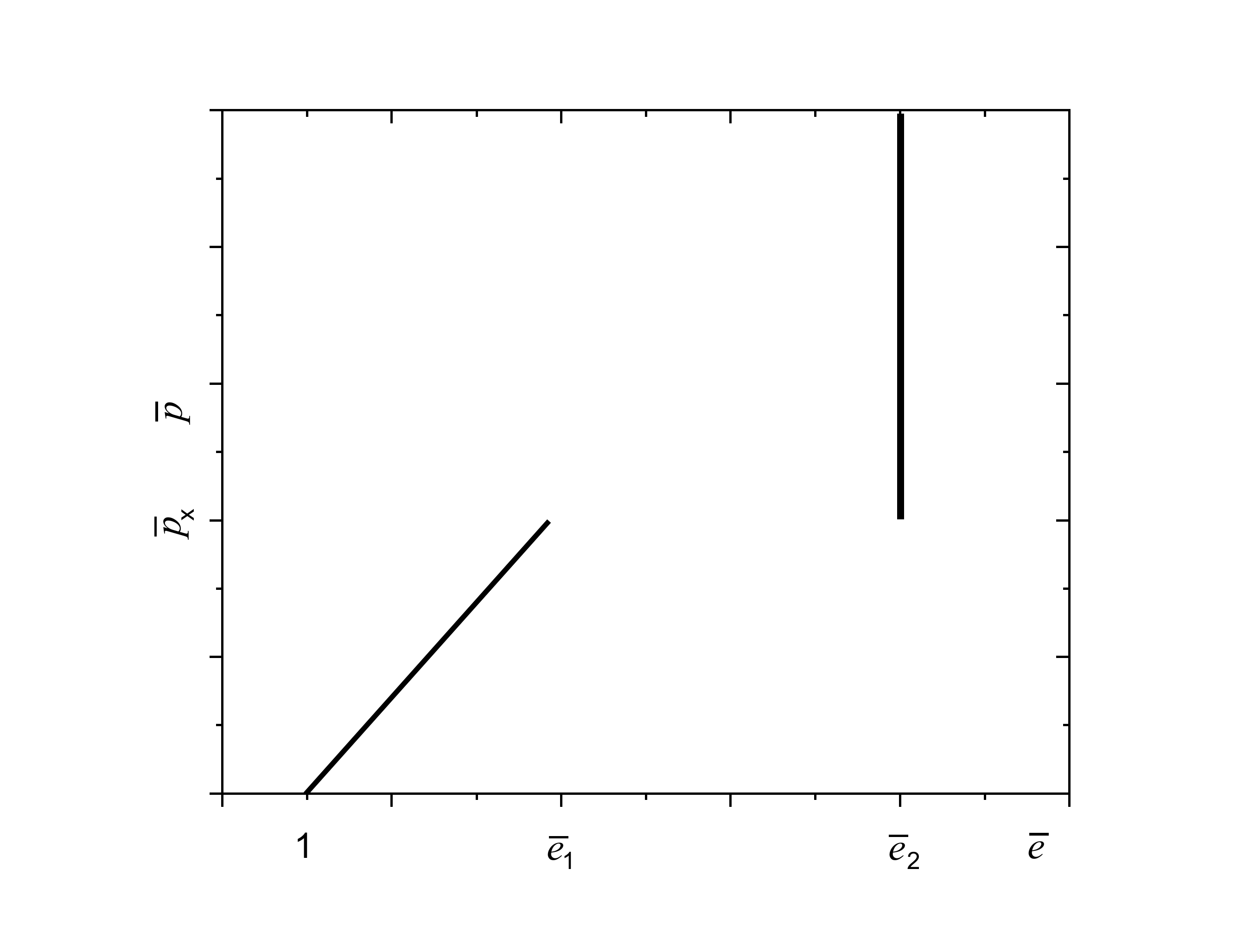}
	\caption{Schematic plots of the EoSs employed in Section \ref{csv} (left panel) and in the current Appendix (right panel).
		Left panel: Eq.~(\ref{eq:csvEoS}) for $v_s^{-2} = 3$ and $\bar p_x = 1$. 
		The asterisk marks the endpoint of the \blau{fiducial SM matter} EoS
		(solid black line, \blau{which is a substitute of a trustable EoS}), resulting in the mass-radius curve exhibited in Fig.~\ref{fig:2} by the solid black curve. 
		Beyond $\bar p = 1$, any EoS is conceivable.
		The dotted line is one possible extension of the \blau{fiducial} EoS, here \blau{especially} with the same sound velocity.
		The resulting mass-radius relation is depicted in Fig.~\ref{fig:2} by the dotted black curve
		for Standard-Model matter only. \blau{Another conceivable continuation could be analog to the pattern displayed in the right panel.}
		Right panel: Eq.~(\ref{EoS}) with $\bar e_0 = 1$ (by definition) and
		$\bar e_1 = \bar e_0 + v_s^{-2} \bar p_x$ and $\bar e_2 = \bar \lambda \bar e_1$.
		Several values of $\bar p_x$ (here the critical pressure of a first-order phase transition)
		are considered, while $v_s^{-2} = 3$ and $\bar \lambda = 3$ are kept fix. 
		The same EoSs apply for SM matter and DM Mirror-World matter.  
		\label{fig:EoSs} 
	}
\end{figure}

\begin{figure}[tb!]
	\includegraphics[width=0.49\columnwidth]{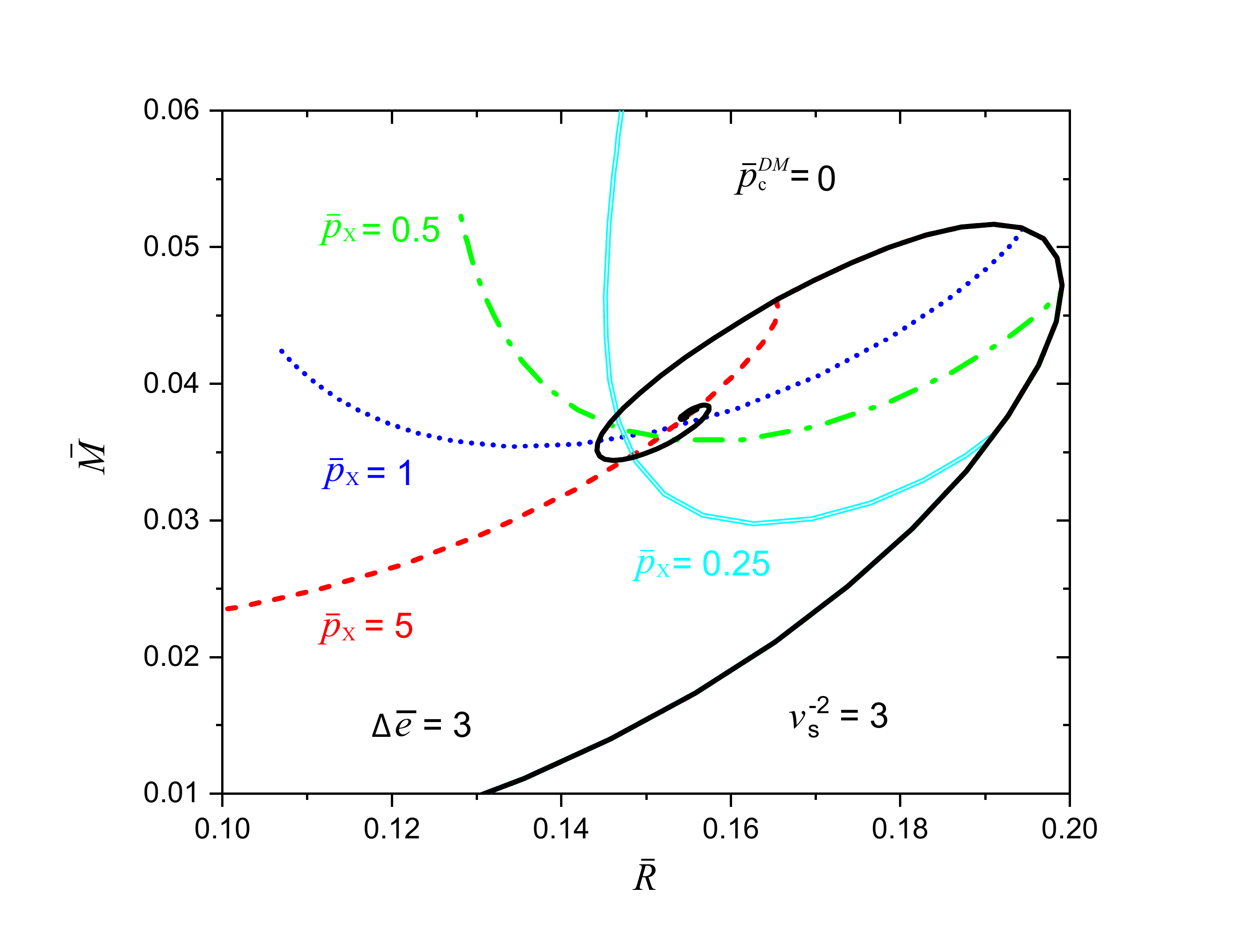} 
	\includegraphics[width=0.49\columnwidth]{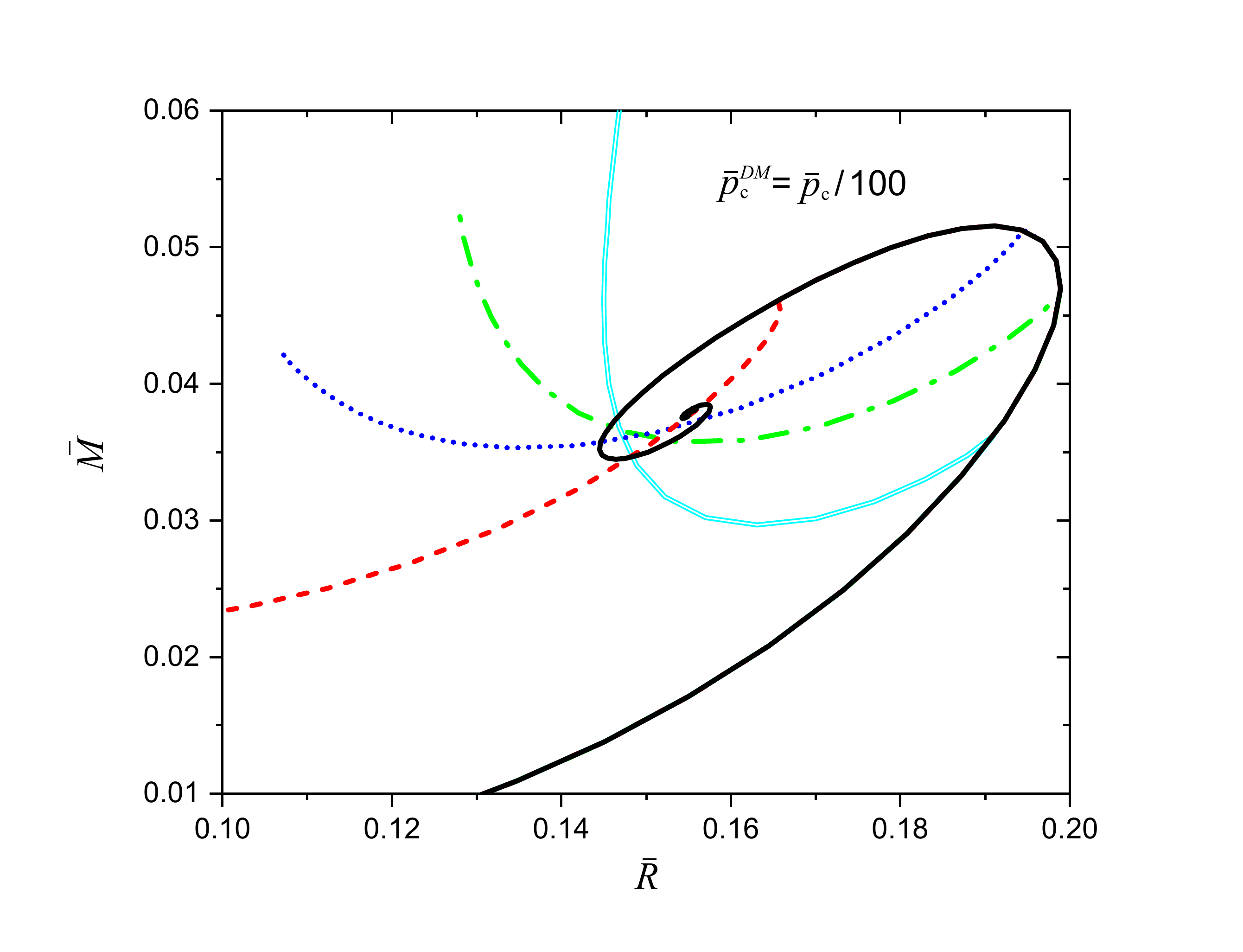} \\[-0.66cm]
	\includegraphics[width=0.49\columnwidth]{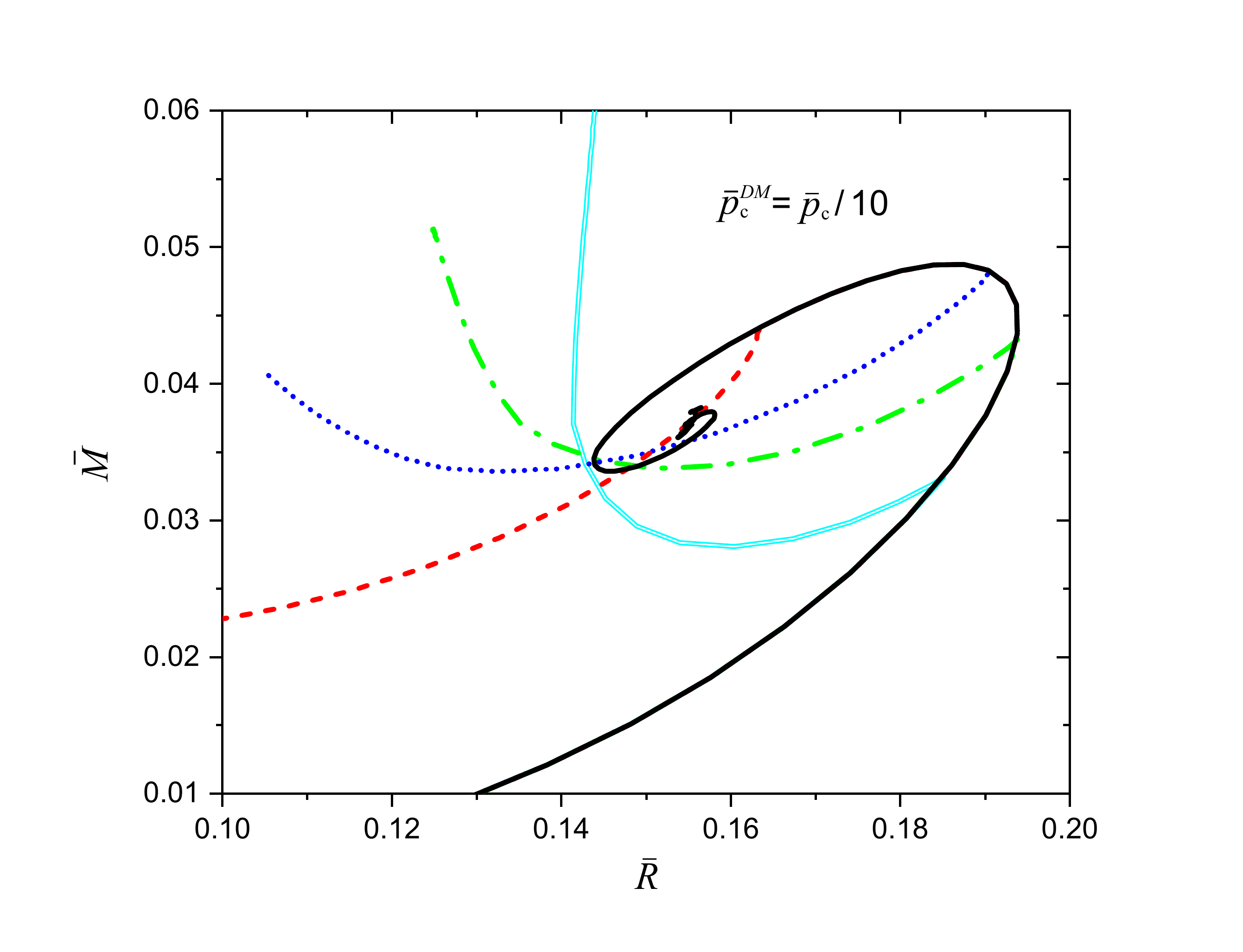}
	\includegraphics[width=0.49\columnwidth]{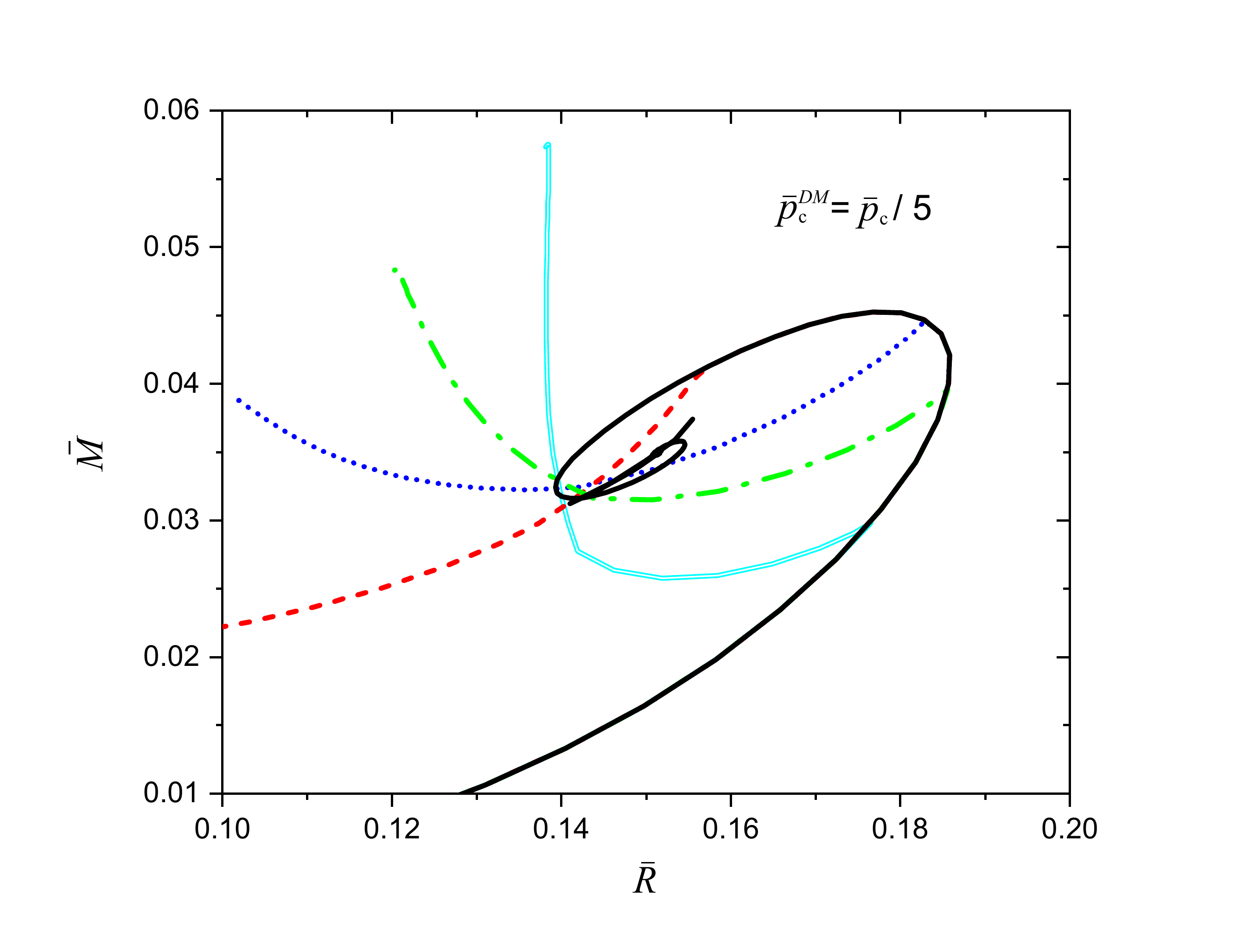} \\[-0.66cm]
	\includegraphics[width=0.49\columnwidth]{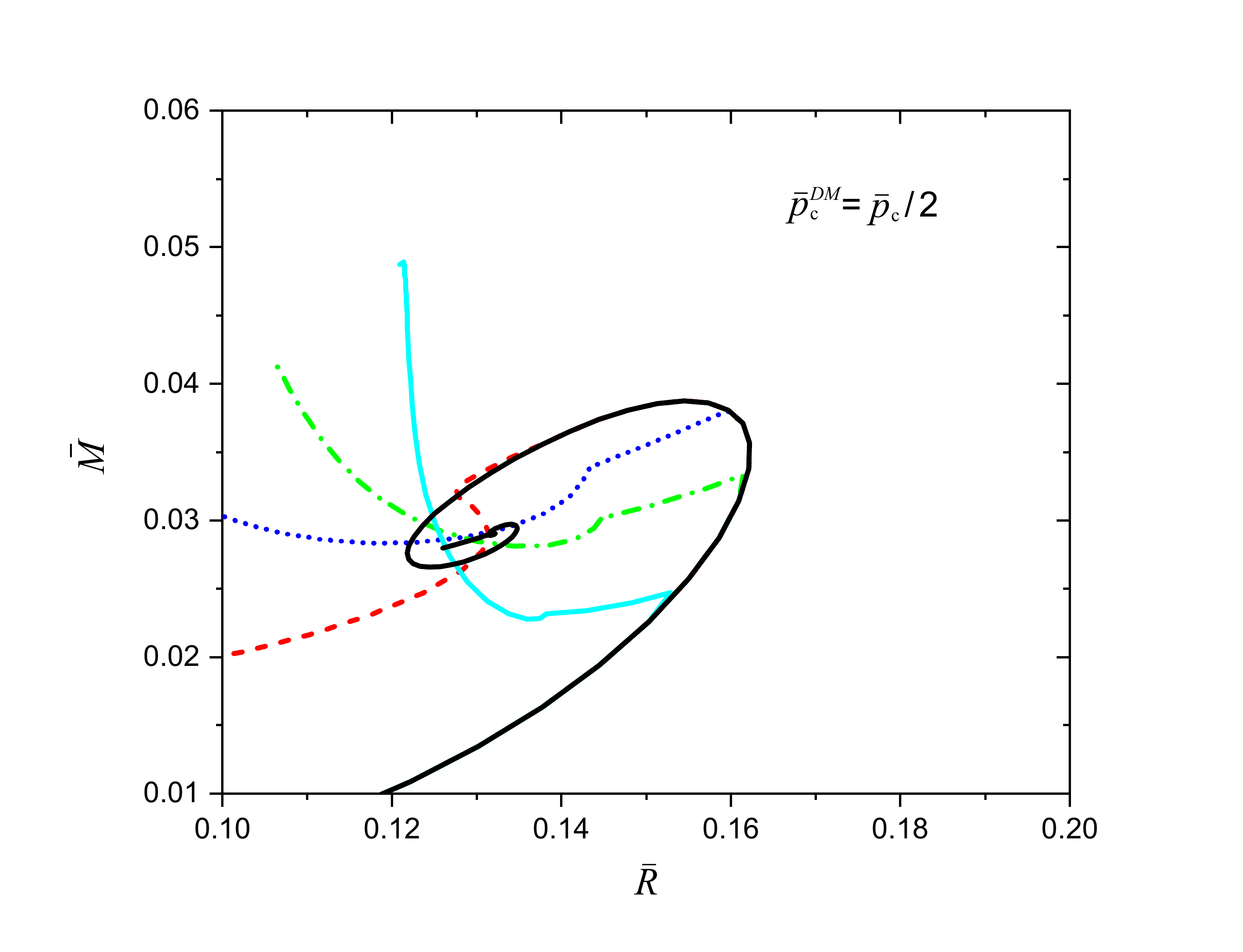}
	\includegraphics[width=0.49\columnwidth]{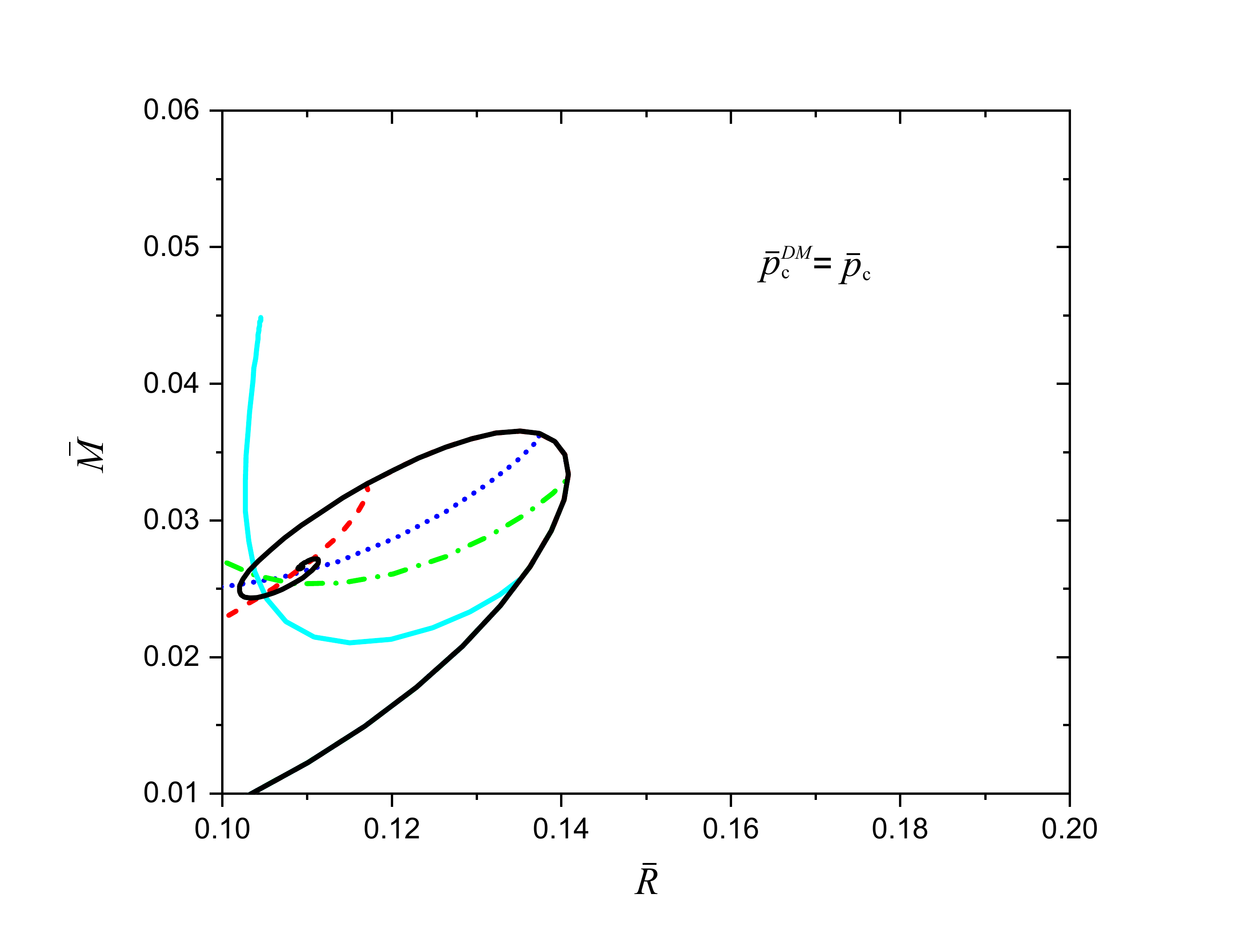}
	\caption{Mass-radius relations of neutron stars with Dark-Matter admixture
		for $v_s^{-2} = 3$ and $\Delta \bar e / \bar e_0 = 3$ and various values of $\bar p_x \equiv p_x /e_0$
		(black solid: $\infty$ [i.e.\ no phase transition], cyan double line: $0.25$, green dot-dashed: $0.5$,
		blue dotted: $1$, red dashed: $5$). 
		The equation of state Eq.~(\ref{EoS}) is used. 
		The DM admixture is determined by the relation $\bar p_c^{DM} = x \bar p_c $
		with $x = 0$ (left top panel, $\bar r_x^{DM} = \bar m_x^{DM} = 0$) 
		to 1 (right bottom panel, $\bar r_x^{DM} = \bar r_x$, $\bar m_x^{DM} = \bar m_x$).
		The central pressures are very large towards the endpoints of curves terminating within the panels. 
		\label{fig:4} 
	}
\end{figure}

Fluid $i = 1$ is attributed to Standard-Model matter (SM) and should represent dense strong-interaction 
(hadron $\cdots$ quark) matter
in $\beta$ equilibrium with local electric charge neutrality. Fluid $i = 2$ stands for a non-self-annihilating Dark-Matter (DM) candidate which we describe also by Eq.~(\ref{EoS}), i.e.\ $\bar p^{(i)} (\bar e^{(i)})$ has the same functional dependence
for all $i$, thus reducing tremendously the parameter space. What distinguishes now SM and DM is the central pressure:
$\bar p_c^{(1)}$ and $\bar p_c^{(2)}$ parameterize the solutions 
$\bar p^{(i)} (\bar r ; \bar p_c^{(1)},  \bar p_c^{(2)})$ and 
$\bar m^{(i)} (\bar r ; \bar p_c^{1},  \bar p_c^{2})$ 
of (\ref{TOV} - \ref{EoS}) yielding, e.g., 
$\bar M = \bar M^{(1)} + \bar M^{(2)}$ as total gravitational star mass 
and visible-matter circumferential radius $\bar R = \bar R^{(1)}$
for choices $\bar p_c^{(2)} = x \bar p_c^{(1)}$ with \blau{$x \in [0,1]$,} facilitating $\bar R^{(2)} \le \bar R^{(1)}$.

One could also think on fluid $i = 2$ as matter in a hidden Mirror World which duplicates our SM world
and communicates essentially only via gravity (cf. \cite{Alizzi:2021vyc,DiLuzio:2021gos,Beradze:2019yyp} 
for recent activities in the field founded fifty years ago \cite{Kobzarev:1966qya}).
Particles of the Mirror World apparently represent candidates of Dark Matter \blau{\cite{Blinnikov:1982eh, Hodges:1993yb} (cf.~\cite{Goldman:2019dbq, Berezhiani:2021src, Berezhiani:2005hv} for further implications and \cite{Berezhiani:2000gw,Berezhiani:2000gh,Berezhiani:2003xm} for cosmological aspects, including SM matter -- DM portal/mixing/feeble-interaction issues)}.
Such a picture justifies the application of the EoS, Eq.~(\ref{EoS}), for both SM matter and DM, see
\cite{Berezhiani:2020zck}. 
We emphasize that, in this Appendix, DM described by Eq.~(\ref{EoS}) is possible at all pressures
$\bar p \gtreqqless \bar p_x$, while in Section \ref{csv} no DM is allowed at $\bar p < \bar p_x$.
To expose this difference we contrast the EoSs used in Section \ref{csv} and in the current Appendix
in Fig.~\ref{fig:EoSs}.

Numerical results, relying on regularity in the center, $\bar m^{(i)} = 0 + {\landau} (\bar r^3)$,
and boundary values $\bar p^{(i)} = \bar p_c^{(i)} -  {\landau} (\bar r^2)$ at $\bar r \to 0$,
are exhibited in Fig.~\ref{fig:4} for $v_s^{-2} = 3$.
(For the dependence on $v_s^2$, see \cite{Schulze:2009dy} [a factor of 
$\sqrt{197.32}$ is to be applied for $\bar M$ and $\bar R$ in Figure~7 there].)
The solid black curves display the mass-radius relation for $\bar p_x \to \infty$, i.e.\
the constant sound velocity EoS is extended unlimited to high pressure (while in Section~\ref{csv}
it was restricted to $p \le p_x$). Reducing gradually the value of $\bar p_x$ means including 
\blau{an} energy density jump from $\bar e_x$ to $\bar \lambda \bar e_x$ at pressure $\bar p_x$
(mimicking a sharp first-order phase transition). The mass-radius curves become strongly modified
(see colored curves),
since the density jump $\Delta \bar e$ is fairly large. In particular, for $\bar p_x \le 1$, an unstable section of the
mass-radius curve is is caused, where the slope is negative, followed by a sequence of stable configurations,
where the slope is positive again, \blau{see the respective figures 2 in \cite{Li:2021sxb, Ranea-Sandoval:2015ldr, Alford:2015gna} for the very pattern of stable-unstable-stable branches}. Equal-mass points on the stable sections represent twin stars.
The case of $\bar p_c^{DM } \equiv \bar p_c^{(i=2)} = 0$ (left top panel, $x = 0$) corresponds to no DM at all, as in Fig.~\ref{fig:2}, and exhibits various cuts provided by an implicit relation of $\bar m_x (\bar r_x)$. Thus, it supplements directly Fig.~\ref{fig:2}. 

With increasing DM fraction, parameterized by $x$, the pattern seen in the left top panel of Fig.~\ref{fig:4}
($x = 0$)
is shifted to left-down, i.e.\ the configurations become more and more compact. 
A tiny admixture of DM, e.g.\ $x = 0.01$ (right top panel), 
does hardly have any noticeable impact on the mass-radius curves.
For $x = 0.1$, one can recognize the left-down shift of the pattern, which becomes somewhat larger at
$x= 0.2$, see middle row. The subtle interplay of $\bar p_c$ and $\bar p_c^{DM}$ and $\bar p_x$
causes additional features at $x \approx 0.5$ due to slightly displaced jumps in the energy density profiles, 
see left bottom panel. The special value 
$x = 1$ means that SM matter and DM have the same pressure and energy density profiles,
i.e.\ $\bar p^{(i=1)} (\bar r) = \bar p^{(i=2)} (\bar r)$ and  $\bar e^{(i=1)} (\bar r) = \bar e^{(i=2)} (\bar r)$ 
implying $\bar m^{(i=1)} (\bar r) = \bar m^{(i=2)} (\bar r)$. The resulting
configurations become most compact, as exhibited in the right bottom panel. 
For $p^{(i)} (r) = p (r)$ and $m^{(i)} (r) = m(r)$, the generalized Schwarzschild pressure profile
in the uniform-density core of $n$ mutually non-interacting fluids, with $e_2$ each, reads
\begin{eqnarray}
	p_{Schwarz, n} (r) &=& - e_2 \frac{(e_2 +p_c) - (e_2 + 3 p_c) W_n }{3(e_2+p_c) - (e_2 + 3 p_c) W_n}, \\
	W_n (r) &:=& \sqrt{1 - n \frac{8 \pi}{3} G_N e_2 r^2},
\end{eqnarray}
where the scaling, e.g.\  by $e_2 = \bar \lambda (e_0 + v_s^{-2} p_x)$ when referring to the EoS (\ref{EoS}),
can be applied.
The core radius follows from $p_{Schwarz, n} (r)  = p_x$; it scales as $\propto 1/\sqrt{n}$.

The selected range of $\bar p_x$ causes a crossing of the mass-radius curves within narrow regions.
This can be considered as reminiscence of the ``special point" discussed in \cite{Cierniak:2021knt,Cierniak:2020eyh}.

When relating these panels to Fig.~\ref{fig:2}, one must ensure that, at radius $\bar r_x$, the DM component must have vanished. With account of that requirement, the panels in Fig.~\ref{fig:4} provide explicit examples of possible cuts in the mass-radius plane of Fig.~\ref{fig:2} \blau{under variation of $\bar p_x$}. One could make this relation more explicitly by digging out the information of $\bar m_x$ and $\bar r_x$ along the mass-radius curves in Fig.~\ref{fig:4}. We leave such a study for follow-up work.

\end{appendix}

%=====================================


\begin{thebibliography}{999}

\bibitem{Pang:2021jta}
P.~T.~H.~Pang, I.~Tews, M.~W.~Coughlin, M.~Bulla, C.~Van Den Broeck and T.~Dietrich,
\textit{Nuclear Physics Multimessenger Astrophysics Constraints on the Neutron Star Equation of State: Adding NICER\textquoteright{}s PSR J0740+6620 Measurement},
Astrophys. J. \textbf{922}, 14 (2021).

\bibitem{Annala:2021gom}
E.~Annala, T.~Gorda, E.~Katerini, A.~Kurkela, J.~N\"attil\"a, V.~Paschalidis and A.~Vuorinen,
\textit{Multimessenger constraints for ultra-dense matter},
\blau{Phys. Rev. X \textbf{12}, 011058 (2022)}.

\bibitem{Yu:2021nvx}
J.~Yu, H.~Song, S.~Ai, H.~Gao, F.~Wang, Y.~Wang, Y.~Lu, W.~Fang and W.~Zhao,
\textit{Multimessenger Detection Rates and Distributions of Binary Neutron Star Mergers and Their Cosmological Implications},
Astrophys. J. \textbf{916}, 54 (2021).

\bibitem{Nicholl:2021rcr}
M.~Nicholl, B.~Margalit, P.~Schmidt, G.~P.~Smith, E.~J.~Ridley and J.~Nuttall,
\textit{Tight multimessenger constraints on the neutron star equation of state from GW170817 and a forward model for kilonova light-curve synthesis},
Mon. Not. Roy. Astron. Soc. \textbf{505}, 3016 (2021).

\bibitem{Margutti:2020xbo}
R.~Margutti and R.~Chornock,
\textit{First Multimessenger Observations of a Neutron Star Merger},
Ann. Rev. Astron. Astrophys. \textbf{59}, 155 (2021).

\bibitem{Tang:2020koz}
S.~P.~Tang, J.~L.~Jiang, W.~H.~Gao, Y.~Z.~Fan and D.~M.~Wei,
\textit{Constraint on phase transition with the multimessenger data of neutron stars},
Phys. Rev. D \textbf{103}, 063026 (2021).

\bibitem{Tews:2020ylw}
I.~Tews, P.~T.~H.~Pang, T.~Dietrich, M.~W.~Coughlin, S.~Antier, M.~Bulla, J.~Heinzel and L.~Issa,
\textit{On the Nature of GW190814 and Its Impact on the Understanding of Supranuclear Matter},
Astrophys. J. Lett. \textbf{908}, L1 (2021).

\bibitem{Silva:2020acr}
H.~O.~Silva, A.~M.~Holgado, A.~C\'ardenas-Avenda\~no and N.~Yunes,
\textit{Astrophysical and theoretical physics implications from multimessenger neutron star observations},
Phys. Rev. Lett. \textbf{126}, 181101 (2021).

\bibitem{Riley:2021pdl}
T.~E.~Riley, A.~L.~Watts, P.~S.~Ray, S.~Bogdanov, S.~Guillot, S.~M.~Morsink, A.~V.~Bilous, Z.~Arzoumanian, D.~Choudhury and J.~S.~Deneva et al.,
\textit{A NICER View of the Massive Pulsar PSR J0740+6620 Informed by Radio Timing and XMM-Newton Spectroscopy},
Astrophys. J. Lett. \textbf{918}, L27 (2021).

\bibitem{Miller:2021qha}
M.~C.~Miller, F.~K.~Lamb, A.~J.~Dittmann, S.~Bogdanov, Z.~Arzoumanian, K.~C.~Gendreau, S.~Guillot, W.~C.~G.~Ho, J.~M.~Lattimer and M.~Loewenstein et al.,
\textit{The Radius of PSR J0740+6620 from NICER and XMM-Newton Data},
Astrophys. J. Lett. \textbf{918}, L28 (2021).

\bibitem{Miller:2019cac}
M.~C.~Miller, F.~K.~Lamb, A.~J.~Dittmann, S.~Bogdanov, Z.~Arzoumanian, K.~C.~Gendreau, S.~Guillot, A.~K.~Harding, W.~C.~G.~Ho and J.~M.~Lattimer et al.,
\textit{PSR J0030+0451 Mass and Radius from $NICER$ Data and Implications for the Properties of Neutron Star Matter},
Astrophys. J. Lett. \textbf{887}, L24 (2019).

\bibitem{Chatziioannou:2020pqz}
K.~Chatziioannou,
\textit{Neutron star tidal deformability and equation of state constraints},
Gen. Rel. Grav. \textbf{52}, 109 (2020).

\bibitem{Christian:2019qer}
J.~E.~Christian and J.~Schaffner-Bielich,
\textit{Twin Stars and the Stiffness of the Nuclear Equation of State: Ruling Out Strong Phase Transitions below $1.7n_0$ with the New NICER Radius Measurements},
Astrophys. J. Lett. \textbf{894}, L8 (2020).

\bibitem{Motta:2022nlj}
T.~F.~Motta and A.~W.~Thomas,
\textit{The role of baryon structure in neutron stars},
Mod. Phys. Lett. A \textbf{37}, 2230001 (2022).

\bibitem{Jokela:2021vwy}
N.~Jokela, M.~J\"arvinen and J.~Remes,
\textit{Holographic QCD in the NICER era},
\blau{Phys. Rev. D \textbf{105}, 086005 (2022)}.

\bibitem{Kovensky:2021kzl}
N.~Kovensky, A.~Poole and A.~Schmitt,
\textit{Building a realistic neutron star from holography},
Phys. Rev. D \textbf{105}, 034022 (2022).

\bibitem{Zhang:2021xdt}
N.~B.~Zhang and B.~A.~Li,
\textit{Impact of NICER\textquoteright{}s Radius Measurement of PSR J0740+6620 on Nuclear Symmetry Energy at Suprasaturation Densities},
Astrophys. J. \textbf{921}, 111 (2021).

\bibitem{Pereira:2020cmv}
J.~P.~Pereira, M.~Bejger, L.~Tonetto, G.~Lugones, P.~Haensel, J.~L.~Zdunik and M.~Sieniawska,
\textit{Probing elastic quark phases in hybrid stars with radius measurements},
Astrophys. J. \textbf{910}, 145 (2021).

\bibitem{Christian:2020xwz}
J.~E.~Christian and J.~Schaffner-Bielich,
\textit{Supermassive Neutron Stars Rule Out Twin Stars},
Phys. Rev. D \textbf{103}, 063042 (2021).

\bibitem{Gerlach:1968zz}
U.~H.~Gerlach,
\textit{Equation of State at Supranuclear Densities and the Existence of a Third Family of Superdense Stars},
Phys. Rev. \textbf{172}, 1325 (1968).

\bibitem{Kampfer:1981yr}
B.~K\"ampfer,
\textit{On the Possibility of Stable Quark and Pion Condensed Stars},
J. Phys. A \textbf{14}, L471 (1981).

\bibitem{Kampfer:1981zmq}
B.~K\"ampfer,
\textit{On stabilizing effects of relativity in cold spheric stars with a phase transition in the interior},
Phys. Lett. B \textbf{101}, 366 (1981).

\bibitem{Li:2019fqe}
J.~J.~Li, A.~Sedrakian and M.~Alford,
\textit{Relativistic hybrid stars with sequential first-order phase transitions and heavy-baryon envelopes},
Phys. Rev. D \textbf{101}, 063022 (2020).

\bibitem{Alford:2017qgh}
M.~G.~Alford and A.~Sedrakian,
\textit{Compact stars with sequential QCD phase transitions},
Phys. Rev. Lett. \textbf{119}, 161104 (2017).

\bibitem{Malfatti:2020onm}
G.~Malfatti, M.~G.~Orsaria, I.~F.~Ranea-Sandoval, G.~A.~Contrera and F.~Weber,
\textit{Delta baryons and diquark formation in the cores of neutron stars},
Phys. Rev. D \textbf{102}, 063008 (2020).

\bibitem{Pereira:2022stw}
J.~P.~Pereira, M.~Bejger, J.~L.~Zdunik and P.~Haensel,
\textit{Differentiating sharp phase transitions from mixed states in neutron stars},
[arXiv:2201.01217 [astro-ph.HE]].

\bibitem{Bejger:2016emu}
M.~Bejger, D.~Blaschke, P.~Haensel, J.~L.~Zdunik and M.~Fortin,
\textit{Consequences of a strong phase transition in the dense matter equation of state for the rotational evolution of neutron stars},
Astron. Astrophys. \textbf{600}, A39 (2017).

\bibitem{Glendenning:1998ag}
N.~K.~Glendenning and C.~Kettner,
\textit{Nonidentical neutron star twins},
Astron. Astrophys. \textbf{353}, L9 (2000).

\bibitem{Jakobus:2020nxw}
P.~Jakobus, A.~Motornenko, R.~O.~Gomes, J.~Steinheimer and H.~Stoecker,
\textit{The possibility of twin star solutions in a model based on lattice QCD thermodynamics},
Eur. Phys. J. C \textbf{81}, 41 (2021).

\bibitem{Alford:2004pf}
M.~Alford, M.~Braby, M.~W.~Paris and S.~Reddy,
\textit{Hybrid stars that masquerade as neutron stars},
Astrophys. J. \textbf{629}, 969 (2005).

\bibitem{Reed:2021nqk}
B.~T.~Reed, F.~J.~Fattoyev, C.~J.~Horowitz and J.~Piekarewicz,
\textit{Implications of PREX-2 on the Equation of State of Neutron-Rich Matter},
Phys. Rev. Lett. \textbf{126}, 172503 (2021).

\bibitem{Annala:2019puf}
E.~Annala, T.~Gorda, A.~Kurkela, J.~N\"attil\"a and A.~Vuorinen,
\textit{Evidence for quark-matter cores in massive neutron stars},
Nature Phys. \textbf{16}, 907 (2020).

\bibitem{Ayriyan:2021prr}
A.~Ayriyan, D.~Blaschke, A.~G.~Grunfeld, D.~Alvarez-Castillo, H.~Grigorian and V.~Abgaryan,
\textit{Bayesian analysis of multimessenger M-R data with interpolated hybrid EoS},
Eur. Phys. J. A \textbf{57}, 318 (2021).

\bibitem{LopeOter:2019pcq}
E.~Lope Oter, A.~Windisch, F.~J.~Llanes-Estrada and M.~Alford,
\textit{nEoS: Neutron Star Equation of State from hadron physics alone},
J. Phys. G \textbf{46}, 084001 (2019).

\bibitem{Lattimer:2015nhk}
J.~M.~Lattimer and M.~Prakash,
\textit{The Equation of State of Hot, Dense Matter and Neutron Stars},
Phys. Rept. \textbf{621}, 127 (2016).

\bibitem{Greif:2020pju}
S.~K.~Greif, K.~Hebeler, J.~M.~Lattimer, C.~J.~Pethick and A.~Schwenk,
\textit{Equation of state constraints from nuclear physics, neutron star masses, and future moment of inertia measurements},
Astrophys. J. \textbf{901}, 155 (2020).

\bibitem{Anzuini:2021lnv}
F.~Anzuini, N.~F.~Bell, G.~Busoni, T.~F.~Motta, S.~Robles, A.~W.~Thomas and M.~Virgato,
\textit{Improved treatment of dark matter capture in neutron stars III: nucleon and exotic targets},
JCAP \textbf{11}, 056 (2021).

\bibitem{Bell:2019pyc}
N.~F.~Bell, G.~Busoni and S.~Robles,
\textit{Capture of Leptophilic Dark Matter in Neutron Stars},
JCAP \textbf{06}, 054 (2019).

\bibitem{Das:2021hnk}
H.~C.~Das, A.~Kumar, B.~Kumar and S.~K.~Patra,
\textit{Dark Matter Effects on the Compact Star Properties},
Galaxies \textbf{10}, 14 (2022).

\bibitem{Das:2021yny}
H.~C.~Das, A.~Kumar and S.~K.~Patra,
\textit{Dark matter admixed neutron star as a possible compact component in the GW190814 merger event},
Phys. Rev. D \textbf{104}, 063028 (2021).

\bibitem{Blaschke:2020qqj}
D.~Blaschke, A.~Ayriyan, D.~E.~Alvarez-Castillo and H.~Grigorian,
\textit{Was GW170817 a Canonical Neutron Star Merger? Bayesian Analysis with a Third Family of Compact Stars},
Universe \textbf{6}, 81 (2020).

\bibitem{Newton:2021yru}
W.~G.~Newton, L.~Balliet, S.~Budimir, G.~Crocombe, B.~Douglas, T.~B.~Head, Z.~Langford, L.~Rivera and J.~Sanford,
\textit{Ensembles of unified crust and core equations of state in a nuclear-multimessenger astrophysics environment},
\blau{Eur. Phys. J. A \textbf{58}, 69 (2022)}.

\bibitem{Huth:2021bsp}
S.~Huth, P.~T.~H.~Pang, I.~Tews, T.~Dietrich, A.~L.~F\`evre, A.~Schwenk, W.~Trautmann, K.~Agarwal, M.~Bulla and M.~W.~Coughlin et al.,
\textit{Constraining Neutron-Star Matter with Microscopic and Macroscopic Collisions},
[arXiv:2107.06229 [nucl-th]].

\bibitem{Raaijmakers:2021uju}
G.~Raaijmakers, S.~K.~Greif, K.~Hebeler, T.~Hinderer, S.~Nissanke, A.~Schwenk, T.~E.~Riley, A.~L.~Watts, J.~M.~Lattimer and W.~C.~G.~Ho,
\textit{Constraints on the Dense Matter Equation of State and Neutron Star Properties from NICER\textquoteright{}s Mass\textendash{}Radius Estimate of PSR J0740+6620 and Multimessenger Observations},
Astrophys. J. Lett. \textbf{918}, L29 (2021).

\bibitem{Raaijmakers:2019qny}
G.~Raaijmakers, T.~E.~Riley, A.~L.~Watts, S.~K.~Greif, S.~M.~Morsink, K.~Hebeler, A.~Schwenk, T.~Hinderer, S.~Nissanke and S.~Guillot et al.,
\textit{A $NICER$ view of PSR J0030+0451: Implications for the dense matter equation of state},
Astrophys. J. Lett. \textbf{887}, L22 (2019).

\bibitem{Raaijmakers:2019dks}
G.~Raaijmakers, S.~K.~Greif, T.~E.~Riley, T.~Hinderer, K.~Hebeler, A.~Schwenk, A.~L.~Watts, S.~Nissanke, S.~Guillot and J.~M.~Lattimer et al.,
\textit{Constraining the dense matter equation of state with joint analysis of NICER and LIGO/Virgo measurements},
Astrophys. J. Lett. \textbf{893}, L21 (2020).

\bibitem{Suleiman:2021hre}
L.~Suleiman, M.~Fortin, J.~L.~Zdunik and P.~Haensel,
\textit{Influence of the crust on the neutron star macrophysical quantities and universal relations},
Phys. Rev. C \textbf{104}, 015801 (2021).

\bibitem{Lattimer:2006xb}
J.~M.~Lattimer and M.~Prakash,
\textit{Neutron Star Observations: Prognosis for Equation of State Constraints},
Phys. Rept. \textbf{442}, 109 (2007).

\bibitem{Klahn:2006ir}
T.~Klahn, D.~Blaschke, S.~Typel, E.~N.~E.~van Dalen, A.~Faessler, C.~Fuchs, T.~Gaitanos, H.~Grigorian, A.~Ho and E.~E.~Kolomeitsev, \textit{et al.}
\textit{Constraints on the high-density nuclear equation of state from the phenomenology of compact stars and heavy-ion collisions},
Phys. Rev. C \textbf{74}, 035802 (2006).

\bibitem{Most:2022wgo}
E.~R.~Most, A.~Motornenko, J.~Steinheimer, V.~Dexheimer, M.~Hanauske, L.~Rezzolla and H.~Stoecker,
\textit{Probing neutron-star matter in the lab: connecting binary mergers to heavy-ion collisions},
[arXiv:2201.13150 [nucl-th]].

\bibitem{HADES:2019auv}
J.~Adamczewski-Musch et al. [HADES],
\textit{Probing dense baryon-rich matter with virtual photons},
Nature Phys. \textbf{15}, 1040 (2019).

\bibitem{Stephanov:1999zu}
M.~A.~Stephanov, K.~Rajagopal and E.~V.~Shuryak,
\textit{Event-by-event fluctuations in heavy ion collisions and the QCD critical point},
Phys. Rev. D \textbf{60}, 114028 (1999).

\bibitem{Karsch:2001cy}
F.~Karsch,
\textit{Lattice QCD at high temperature and density},
Lect. Notes Phys. \textbf{583}, 209 (2002).

\bibitem{Fukushima:2010bq}
K.~Fukushima and T.~Hatsuda,
\textit{The phase diagram of dense QCD},
Rept. Prog. Phys. \textbf{74}, 014001 (2011).

\bibitem{Halasz:1998qr}
A.~M.~Halasz, A.~D.~Jackson, R.~E.~Shrock, M.~A.~Stephanov and J.~J.~M.~Verbaarschot,
\textit{On the phase diagram of QCD},
Phys. Rev. D \textbf{58}, 096007 (1998).

\bibitem{Blacker:2020nlq}
S.~Blacker, N.~U.~F.~Bastian, A.~Bauswein, D.~B.~Blaschke, T.~Fischer, M.~Oertel, T.~Soultanis and S.~Typel,
\textit{Constraining the onset density of the hadron-quark phase transition with gravitational-wave observations},
Phys. Rev. D \textbf{102}, 123023 (2020).

\bibitem{Blaschke:2020vuy}
D.~Blaschke and M.~Cierniak,
\textit{Studying the onset of deconfinement with multi-messenger astronomy of neutron stars},
Astron. Nachr. \textbf{342}, 227 (2021).

\bibitem{Orsaria:2019ftf}
M.~G.~Orsaria, G.~Malfatti, M.~Mariani, I.~F.~Ranea-Sandoval, F.~Garc\'\i{}a, W.~M.~Spinella, G.~A.~Contrera, G.~Lugones and F.~Weber,
\textit{Phase transitions in neutron stars and their links to gravitational waves},
J. Phys. G \textbf{46}, 073002 (2019).

\bibitem{Cierniak:2021knt}
M.~Cierniak and D.~Blaschke,
\textit{Hybrid neutron stars in the mass-radius diagram},
Astron. Nachr. \textbf{342}, 819 (2021).

\bibitem{Kampfer:1985mre}
B.~K\"ampfer,
\textit{Phase transitions in dense nuclear matter and explosive neutron star phenomena},
Phys. Lett. B \textbf{153}, 121 (1985).

\bibitem{Kampfer:1983we}
B.~K\"ampfer,
\textit{Phase transitions in nuclear matter and consequences for neutron stars}, 
J. Phys. G \textbf{9}, 1487 (1983).

\bibitem{Schertler:2000xq}
K.~Schertler, C.~Greiner, J.~Schaffner-Bielich and M.~H.~Thoma,
\textit{Quark phases in neutron stars and a 'third family' of compact stars as a signature for phase transitions},
Nucl. Phys. A \textbf{677}, 463 (2000).

\bibitem{Christian:2017jni}
J.~E.~Christian, A.~Zacchi and J.~Schaffner-Bielich,
\textit{Classifications of Twin Star Solutions for a Constant Speed of Sound Parameterized Equation of State},
Eur. Phys. J. A \textbf{54}, 28 (2018).

\bibitem{Migdal:1979je}
A.~B.~Migdal, A.~I.~Chernoutsan and I.~N.~Mishustin,
\textit{Pion condensation and dynamics of neutron stars},
Phys. Lett. B \textbf{83}, 158-160 (1979).

\bibitem{Kampfer:1983zz}
B.~K\"ampfer,
\textit{On the collapse of neutron stars and stellar cores to pion-condensed stars},
Astrophys. Space Sci. \textbf{93}, 185-197 (1983).

\bibitem{Haensel:1987}
J.~L.~Zdunik, P.~Haensel and R.~Schaeffer,
\textit{Phase transitons in stellar cores, II Equilibrium configurations in general relativity},
Astron.\ Astrophys.\ \textbf{172}, 95 (1987).

\bibitem{Schaffner-Bielich:2020psc}
J.~Schaffner-Bielich,
\textit{Compact Star Physics},
Cambridge University Press (2020).

\bibitem{Buchdahl}
H.~A.~Buchdahl, \textit{General-relativistic fluid spheres. III. A static gaseous model},
Astrophys.\ J.\ \textbf{147}, 310 (1967).

\bibitem{Lattimer:2000nx}
J.~M.~Lattimer and M.~Prakash,
\textit{Neutron star structure and the equation of state},
Astrophys. J. \textbf{550}, 426 (2001).

\bibitem{Dengler:2021qcq}
Y.~Dengler, J.~Schaffner-Bielich and L.~Tolos,
\textit{Second Love number of dark compact planets and neutron stars with dark matter},
Phys. Rev. D \textbf{105}, 043013 (2022).

\bibitem{Karkevandi:2021ygv}
\blau{D.~R.~Karkevandi, S.~Shakeri, V.~Sagun and O.~Ivanytskyi,
\textit{Bosonic dark matter in neutron stars and its effect on gravitational wave signal},
Phys. Rev. D \textbf{105}, 023001 (2022).}

\bibitem{Haensel:1983}
J.~L.~Zdunik, P.~Haensel and R.~Schaeffer,
\textit{Phase transitons in stellar cores, I Equilibrium configurations},
Astron.\ Astrophys.\ \textbf{126}, 121 (1983).

\bibitem{Alizzi:2021vyc}
A.~Alizzi and Z.~K.~Silagadze,
\textit{Dark photon portal into mirror world},
Mod. Phys. Lett. A \textbf{36}, 2150215 (2021).

\bibitem{DiLuzio:2021gos}
L.~Di Luzio, B.~Gavela, P.~Quilez and A.~Ringwald,
\textit{Dark matter from an even lighter QCD axion: trapped misalignment},
JCAP \textbf{10}, 001 (2021).

\bibitem{Beradze:2019yyp}
R.~Beradze, M.~Gogberashvili and A.~S.~Sakharov,
\textit{Binary Neutron Star Mergers with Missing Electromagnetic Counterparts as Manifestations of Mirror World},
Phys. Lett. B \textbf{804}, 135402 (2020).

\bibitem{Kobzarev:1966qya}
I.~Y.~Kobzarev, L.~B.~Okun and I.~Y.~Pomeranchuk,
\textit{On the possibility of experimental observation of mirror particles},
Sov. J. Nucl. Phys. \textbf{3}, 837 (1966).

\bibitem{Blinnikov:1982eh}
\blau{S.~I.~Blinnikov and M.~Y.~Khlopov,
\textit{On the possible effects of "mirror" particles},
Sov. J. Nucl. Phys. \textbf{36}, 472 (1982).}

\bibitem{Hodges:1993yb}
H.~M.~Hodges,
\textit{Mirror baryons as the dark matter},
Phys. Rev. D \textbf{47}, 456 (1993).

\bibitem{Goldman:2019dbq}
\blau{I.~Goldman, R.~N.~Mohapatra and S.~Nussinov,
\textit{Bounds on neutron-mirror neutron mixing from pulsar timing},
Phys. Rev. D \textbf{100}, 123021 (2019).}

\bibitem{Berezhiani:2021src}
\blau{Z.~Berezhiani,
\textit{Antistars or antimatter cores in mirror neutron stars?},
[arXiv:2106.11203 [astro-ph.HE]] (2021).}

\bibitem{Berezhiani:2005hv}
\blau{Z.~Berezhiani and L.~Bento,
\textit{Neutron - mirror neutron oscillations: How fast might they be?},
Phys. Rev. Lett. \textbf{96}, 081801 (2006).}
	
\bibitem{Berezhiani:2000gw}
\blau{Z.~Berezhiani, D.~Comelli and F.~L.~Villante,
\textit{The Early mirror universe: Inflation, baryogenesis, nucleosynthesis and dark matter},
Phys. Lett. B \textbf{503}, 362-375 (2001).}

\bibitem{Berezhiani:2000gh}
\blau{Z.~Berezhiani, L.~Gianfagna and M.~Giannotti,
\textit{Strong CP problem and mirror world: The Weinberg-Wilczek axion revisited},
Phys. Lett. B \textbf{500}, 286-296 (2001).}

\bibitem{Berezhiani:2003xm}
\blau{Z.~Berezhiani,
\textit{Mirror world and its cosmological consequences},
Int. J. Mod. Phys. A \textbf{19}, 3775-3806 (2004).}

\bibitem{Berezhiani:2020zck}
Z.~Berezhiani, R.~Biondi, M.~Mannarelli and F.~Tonelli,
\textit{Neutron - mirror neutron mixing and neutron stars},
\blau{Eur. Phys. J. C \textbf{81}, 1036 (2022)}.

\bibitem{Schulze:2009dy}
R.~Schulze and B.~K\"ampfer,
\textit{Cold quark stars from hot lattice QCD},
[arXiv:0912.2827 [nucl-th]].

\bibitem{Li:2021sxb}
\blau{J.~J.~Li, A.~Sedrakian and M.~Alford,
\textit{Relativistic hybrid stars in light of the NICER PSR J0740+6620 radius measurement},
Phys. Rev. D \textbf{104}, L121302 (2021).}

\bibitem{Ranea-Sandoval:2015ldr}
\blau{I.~F.~Ranea-Sandoval, S.~Han, M.~G.~Orsaria, G.~A.~Contrera, F.~Weber and M.~G.~Alford,
\textit{Constant-sound-speed parametrization for Nambu\textendash{}Jona-Lasinio models of quark matter in hybrid stars},
Phys. Rev. C \textbf{93}, 045812 (2016).}

\bibitem{Alford:2015gna}
\blau{M.~G.~Alford and S.~Han,
\textit{Characteristics of hybrid compact stars with a sharp hadron-quark interface},
Eur. Phys. J. A \textbf{52}, 62 (2016).}

\bibitem{Cierniak:2020eyh}
M.~Cierniak and D.~Blaschke,
\textit{The special point on the hybrid star mass\textendash{}radius diagram and its multi\textendash{}messenger implications},
Eur. Phys. J. ST \textbf{229}, 3663 (2020).


\end{thebibliography}
\end{document}